\documentclass[conference]{IEEEtran}
\IEEEoverridecommandlockouts

\usepackage{amsmath,amsfonts}
\usepackage{algorithmic}
\usepackage{algorithm}
\usepackage{array}
\usepackage[caption=false,font=normalsize,labelfont=sf,textfont=sf]{subfig}
\usepackage{textcomp}
\usepackage{stfloats}
\usepackage{url}
\usepackage{verbatim}
\usepackage{graphicx}
\usepackage{cite}
\usepackage{color}
\usepackage{enumitem}
\usepackage{amssymb}
\newtheorem{definition}{Definition}

\newtheorem{proposition}[definition]{Proposition}
\newtheorem{lemma}[definition]{Lemma}
\newtheorem{theorem}[definition]{Theorem}
\newtheorem{corollary}[definition]{Corollary}
\newtheorem{example}[definition]{Example}
\newtheorem{remark}[definition]{Remark}
\newcommand {\beq}[1]{\begin{equation}
                      \label{eq:#1} }
\newcommand {\eeq}{\end{equation}}

\newcommand {\req}[1]{(\ref{eq:#1})}

\newcommand {\bear}[1]{
                       \begin{eqnarray}
                       \label{eq:#1} }
\newcommand {\eear}{\end{eqnarray}}

\newcommand {\bearn}{\begin{eqnarray*}}
\newcommand {\eearn}{\end{eqnarray*}}
\newcommand {\bsec}[2]{
                       \section{#1}
                       \label{sec:#2} }

\newcommand {\rsec}[1]{Section \ref{sec:#1}}
\newcommand {\bsubsec}[2]{
                       \subsection{#1}
                       \label{sec:#2} }
\newcommand {\bsubsubsec}[2]{
                       \subsubsection{#1}
                       \label{sec:#2} }

\newcommand {\rfig}[1]{Figure \ref{fig:#1}}

\newcommand {\bdefin}[1]{\begin{definition}
                             \label{def:#1} }
\newcommand {\edefin}       {\end{definition}}
\newcommand {\rdef}[1]{Definition \ref{def:#1}}
\newcommand {\bprop}[1]{\begin{proposition}
                              \label{prop:#1} }
\newcommand {\eprop}       {\end{proposition}}

\newcommand {\blem}[1]{\begin{lemma}
                            \label{lem:#1} }
\newcommand {\elem}   {\end{lemma}}
\newcommand {\rlem}[1]{Lemma \ref{lem:#1}}

\newcommand {\bthe}[1]{\begin{theorem}
                         \label{the:#1} }
\newcommand {\ethe}   {\end{theorem}}
\newcommand {\rthe}[1]{Theorem \ref{the:#1}}

\newcommand {\bproof}{\noindent {\bf Proof.} \ }
\newcommand {\eproof} {\hfill $\square$ \\ \vspace{.3cm}}
\newcommand {\bcor}[1]{\begin{corollary}
                           \label{cor:#1} }
\newcommand {\ecor}   {\end{corollary}}
\newcommand {\rcor}[1]{Corollary \ref{cor:#1}}
\newcommand {\bex}[2]{\vspace{.1in}
                      \begin{example}
                                {\bf #2}
                      \label{ex:#1} }
\newcommand {\eex}       {\end{example} \vspace{.3cm} }
\newcommand {\rex}[1]{Example \ref{ex:#1}}
\newcommand {\brem}[1]{\begin{remark}
                       \label{rem:#1} \em }
\newcommand {\erem}   {\end{remark}}

\def\ex{{\bf\sf E}}
\def\pr{{\bf\sf P}}

\newcommand{\gen}{g}

\begin{document}

\title{Consistent Channel Hopping Algorithms for the Multichannel Rendezvous Problem with Heterogeneous Available Channel Sets}

\author{Yiwei Liu,
        Yi-Chia Cheng, and
		Cheng-Shang~Chang,~\IEEEmembership{Fellow,~IEEE,}
		\IEEEcompsocitemizethanks{\IEEEcompsocthanksitem
		The authors are with the Institute of Communications Engineering, National Tsing Hua University, Hsinchu 300044, Taiwan R.O.C. Email: s112064521@m112.nthu.edu.tw; s111064701@m111.nthu.edu.tw;  cschang@ee.nthu.edu.tw.
		\protect\\
	}
\thanks{
Part of this work was presented in 2025 IEEE Wireless Communications and Networking Conference (WCNC) \cite{CLC25}. This work was supported in part by the Ministry of Science and Technology, Taiwan, under Grant 111-2221-E-007-038-MY3 and 111-2221-E-007-045-MY3. (Corresponding author: Cheng-Shang Chang.)}}

\maketitle

\begin{abstract}
We propose a theoretical framework for \emph{consistent channel hopping algorithms} to address the multichannel rendezvous problem (MRP) in wireless networks with heterogeneous available channel sets. A channel selection function is called consistent if the selected channel remains unchanged when the available channel set shrinks, provided the selected channel is still available. We show that all consistent channel selection functions are equivalent to the function that always selects the smallest-index channel under appropriate channel relabeling. This leads to a natural representation of a consistent channel hopping algorithm as a sequence of permutations. For the two-user MRP, we characterize rendezvous time slots using a fictitious user and derive tight bounds on the maximum time-to-rendezvous (MTTR) and expected time-to-rendezvous (ETTR). Notably, the ETTR is shown to be the inverse of the Jaccard index when permutations are randomly selected. We also prove that consistent channel hopping algorithms maximize the rendezvous probability. To reduce implementation complexity, we propose the \emph{modulo algorithm}, which uses modular arithmetic with one-cycle permutations and achieves performance comparable to locality-sensitive hashing (LSH)-based algorithms. The framework is extended to multiple users, with novel strategies such as stick-together, spread-out, and a hybrid method that accelerates rendezvous in both synchronous and asynchronous settings. Simulation results confirm the effectiveness and scalability of the proposed algorithms.
\end{abstract}


\section{Introduction}
\label{sec:introudction}

In  the Internet of Things (IoT), devices operate across multiple communication channels and must discover each other before initiating communication. Since primary users may occupy some channels, secondary users need to find spectrum holes to identify available channels. The multichannel rendezvous problem (MRP) requires that two secondary users find a common channel by hopping across their available channels over time. This problem poses a significant challenge for neighbor discovery in many IoT applications, as discussed in studies such as \cite{Theis2011,Bian2013}.

The MRP has received increasing attention in recent years, as evidenced by several studies and a dedicated monograph on the topic \cite{Book}. As discussed in \cite{GAP2019}, the MRP fundamentally consists of three components: (i) users, (ii) time, and (iii) channels. Early studies (e.g., \cite{SynMAC,Quorum,wjliao,ETCH2013,MOR2015,ToN2015,ChangGC17,CRISS2019}) primarily focused on a relatively simple setting characterized by the following assumptions: (i) symmetric users (sym), where users are indistinguishable; (ii) synchronized clocks (sync); (iii) homogeneous available channels (homo), meaning all users have identical channel sets; and (iv) globally labeled channels (global), where each user's channel labels are consistent with global identifiers. This configuration is referred to as the sym/sync/homo/global MRP in \cite{GAP2019}.

In real-world scenarios, however, clocks are typically not synchronized at startup, prompting interest in the sym/async/homo/global setting. To address this, researchers have developed channel hopping (CH) sequences based on combinatorial designs. Notable examples include CRSEQ \cite{CRSEQ}, JS \cite{JS2011}, DRDS \cite{DRDS13}, T-CH \cite{Matrix2015}, DSCR \cite{DSCR2016}, IDEAL-CH \cite{GAP2019}, and RDSML-CH \cite{Wang2022}. For an MRP with \( N \) total channels, these CH sequences ensure rendezvous within \( O(N^2) \) time slots.

A more difficult variant arises when users do not share the same available channels---a situation denoted as the sym/async/hetero/global MRP \cite{GAP2019}. In this context, the focus shifts toward minimizing the maximum time-to-rendezvous (MTTR). Several state-of-the-art algorithms achieve an MTTR of \( O((\log\log N)n_1 n_2) \), where \( n_1 \) and \( n_2 \) denote the number of available channels for users 1 and 2, respectively \cite{Chen14,Improved2015,Chang18,gu2020heterogeneous}. However, these algorithms often yield worse expected time-to-rendezvous (ETTR) compared to the random algorithm. Two notable exceptions are the quasi-random (QR) algorithm \cite{Quasi2018}, which uses 4B5B encoding, and the two-prime clock method in \cite{ToN2017}. Both techniques achieve ETTRs close to the random algorithm, while maintaining an MTTR of \( O((\log N)n_1 n_2) \). The QECH algorithm \cite{QECH} also offers similar MTTR guarantees.

Jiang and Chang recently introduced a novel approach for the sym/async/hetero/global MRP by leveraging locality-sensitive hashing (LSH) \cite{LSH}. Their method exploits the observation that users are often physically close and thus have similar available channel sets, yielding a high Jaccard index. The Jaccard index \( J \) is defined as
\[
J = \frac{n_{1,2}}{n_1 + n_2 - n_{1,2}},
\]
where \( n_{1,2} \) is the number of common channels between the two users. The authors demonstrated that the ETTR of their LSH-based algorithms can outperform that of the random algorithm when \( J \) is large.
A key innovation in their design is the use of \emph{coupled} CH sequences based on globally labeled channels. This coupling enables the application of the ``focal'' strategy from \cite{Alpern95}, significantly improving rendezvous efficiency.

Motivated by the success of LSH algorithms, we seek fundamental theories behind these algorithms.
For this, we consider a general framework of channel hopping algorithms. We view a channel hopping algorithm as a sequence of channel selection functions $\{\phi_1, \phi_2, \ldots\}$.
Given an available channel set ${\bf c}$ for a user,
the channel selection function $\phi_t$ is used to select a channel in ${\bf c}$ at time $t$.
We identify a set of channel selection functions that possesses the following {\em consistent} property:

 {\em When some channels are removed from the available channel set, and the previously selected channel remains in the set after the removal, the selection stays the same. The selection only needs to be changed if the previously selected channel is no longer in the available channel set after the removal.}

A channel hopping algorithm is called {\em consistent} if all the channel selection functions $\{\phi_1, \phi_2, \ldots\}$ are consistent.

In this paper, we develop a comprehensive theoretical framework for \emph{consistent channel hopping algorithms} to address the MRP, particularly under the challenging setting of heterogeneous available channels. The main contributions of the paper are summarized as follows:

\begin{enumerate}
    \item \textbf{Equivalent Representation:}
    It is shown that every consistent channel selection function is equivalent, under relabeling, to selecting the smallest-index channel. As a result, every consistent channel hopping algorithm can be represented as a sequence of permutations \( \{ \pi_1, \pi_2, \ldots \} \) such that \( \phi_t = \pi_t^{-1} \circ \phi^{\min} \circ \pi_t \), where $\phi^{\min}$ is the channel selection function that selects the smallest-index channel.

    \item \textbf{Characterization of Rendezvous Time Slots:}
    Using a fictitious user whose available channel set is the union of all participating users, we provide a complete characterization of the rendezvous time slots under consistent channel hopping algorithms in the synchronous setting.

    \item \textbf{Tight MTTR Bound:}
    When the permutation sequence is generated by a one-cycle permutation, the maximum time-to-rendezvous (MTTR) is shown to be bounded by \( N - n_{1,2} + 1 \), where \( n_{1,2} \) is the number of common channels.

    \item \textbf{ETTR as Inverse of Jaccard Index:}
    When the permutations are drawn randomly and independently, the expected time-to-rendezvous (ETTR) is proved to be the inverse of the Jaccard index \( J \), i.e., \( \text{ETTR} = 1/J \).

    \item \textbf{Optimality:}
    The rendezvous probability under any channel hopping algorithm is shown to be upper-bounded by \( J \), and consistent algorithms achieve this bound.

    \item \textbf{Modulo Algorithm:}
    A low-complexity modulo-based consistent channel hopping algorithm is proposed, which avoids the need for pseudo-random permutation generation and achieves comparable performance to LSH-based algorithms with only \( O(n) \) complexity.

    \item \textbf{Extension to Multiple Users:}
    The framework is extended to the case of \( K \) users. The paper introduces the stick-together and spread-out strategies, and derives closed-form ETTR expressions using Markov chain analysis when $K=3$.

    \item \textbf{Hybrid Algorithm:}
    A hybrid strategy that alternates between the stick-together and spread-out strategies is proposed, balancing exploration and convergence. It is particularly effective at scale and in asynchronous settings.

    \item \textbf{Asynchronous Setting via Dimension Reduction:}
    The paper generalizes the LSH4 dimension reduction technique \cite{LSH} to the asynchronous setting. A seeded pseudo-random generator ensures deterministic behavior and eventual clock synchronization after rendezvous.

    \item \textbf{Simulation Results:}
    Extensive simulations are conducted for two-user, three-user, and large-scale settings. The results validate the theoretical claims and demonstrate the superior performance of consistent algorithms over existing methods, especially when the Jaccard index is large.
\end{enumerate}

The remainder of the paper is organized as follows. In \rsec{mrp}, we provide a brief overview of the multichannel rendezvous problem. \rsec{theory} presents the theoretical foundations of consistent channel hopping algorithms. In \rsec{exp}, we introduce the modulo algorithm as a low-complexity implementation of consistent channel hopping algorithms. \rsec{musers} discusses the extension of these algorithms to the multichannel rendezvous problem with multiple users. In \rsec{asynset}, we further extend the algorithms to operate in asynchronous settings. Simulation results comparing our proposed algorithms with the random and LSH-based algorithms from \cite{LSH} are presented in \rsec{sim}. Finally, we conclude the paper in \rsec{con}.

\bsec{The multichannel rendezvous problem}{mrp}

The MRP in a wireless network is commonly referred to as the problem of two users (IoT devices) rendezvousing on a commonly available channel by hopping over their available channels over time. A common assumption in the literature is that there is a global channel enumeration system, i.e., channels are globally labeled from 1 to $N$, where $N$ is the total number of channels in a wireless network.

We represent the available channel set for user $k$ (where $k=1, 2$) as
$$
{\bf c}_k = \{ c_{k,1}, \ldots, c_{k,n_k}\},
$$
where $n_k = |{\bf c}_k|$ is the number of available channels for user $k$. We assume that there is at least one channel that is commonly available to the two users (as otherwise, it is impossible for the two users to rendezvous), i.e.,
\beq{avail1111}
{\bf c}_1 \cap {\bf c}_2 \ne \varnothing.
\eeq
Let $n_{1,2} = |{\bf c}_1 \cap {\bf c}_2 |$ be the number of common channels between these two users.

As in the MRP literature, we consider the discrete-time setting, where time is divided into time slots indexed from $t =  1, 2, \ldots$. The time-to-rendezvous (TTR) is defined as the number of time slots needed for two users to hop to a common channel. A rendezvous time slot is when the two users rendezvous.

\bsec{Theoretical analysis}{theory}

In this section, we develop the associated theory for consistent channel hopping algorithms.
 We consider the MRP in the synchronous setting, i.e., the clocks of the two users are synchronized.
The application of consistent channel hopping algorithms in the asynchronous setting will be addressed in \rsec{asynset}.

\bsubsec{Consistent channel selection functions}{ccsf}

We introduce the notion of {\em consistent channel selection functions} and develop their associated mathematical properties.

\bdefin{ccsf0}
Given an available channel set ${\bf c}$ (which is a nonempty subset of $\{1,2, \ldots, N\}$), a channel  selection  function $\phi$ selects  a channel
from ${\bf c}$, i.e.,  $\phi({\bf c})=c$ for some $c \in {\bf c}$. A channel hopping algorithm is characterized by a sequence of channel selection functions
$\{\phi_1, \phi_2, \ldots\}$, where $\phi_t$ is used to select the channel at time $t$ from an available channel set ${\bf c}$.
\edefin

We note that the total number of channel selection functions
is $\prod_{k=2}^{N} k^{\binom{N}{k}}$.
To see this, note that for $|{\bf c}|=1$, there is only one way to define the channel selection function.
For each subset \( {\bf c} \) with \( |{\bf c}| = k \geq 2 \), there are \( k \) choices for \( \phi({\bf c}) \) (since \( \phi({\bf c}) \) can be any element of \( {\bf c} \)). As there are \( {\binom{N}{k}} \) subsets of size \( k \),
there are $k^{\binom{N}{k}}$ choices.

In this paper, we focus on a subset of channel selection functions, called the {\em consistent} channel selection functions.

\bdefin{ccsf}
A channel  selection  function $\phi$ is called {\em consistent} if, for any two available channel sets ${\bf c}^\prime \subset {\bf c}^{\prime\prime}$,  $\phi({\bf c}^{\prime\prime})$ is in  ${\bf c}^\prime$,
the following condition holds:
\beq{consist1111}
\phi({\bf c}^{\prime\prime})=\phi({\bf c}^\prime).
\eeq
\edefin

To understand the insight behind the consistency property in \req{consist1111}, consider the scenario where an available channel $c$ in ${\bf c}^{\prime\prime}$ is suddenly occupied by a primary user and the available channel set is reduced to ${\bf c}^\prime={\bf c}^{\prime\prime} \backslash \{c\}$. If the channel selected by  a consistent channel selection function  is still in ${\bf c}^\prime$, then it continues to select the same channel (without making any change). Only when the selected channel is $c$ does it need to change its selection.

\bex{small}{}
Consider the channel selection function $\phi$ that selects the channel with the {\em smallest} index from the available channel set.
Suppose that $\phi({\bf c}^{\prime\prime})$ belongs to ${\bf c}^\prime$. Since the channel with the smallest index is selected from
${\bf c}^{\prime\prime}$ and ${\bf c}^\prime \subset {\bf c}^{\prime\prime}$,  that same channel will also have the smallest index in ${\bf c}^\prime$, and it is therefore selected from
 ${\bf c}^\prime$.
Thus, the condition in \req{consist1111} holds and
$\phi$ is  a consistent channel selection function. We denote such a channel selection function as $\phi^{\min}$.

Using the same argument, one can show that selecting the channel with the {\em largest} index  from the available channel set is also
a consistent channel selection function (denoted by $\phi^{\max}$).
Furthermore, one can extend this to injective functions (one-to-one functions).
Recall that  a function \( f: A \to B \) is injective if for all \( x_1, x_2 \in A \), \( f(x_1) = f(x_2) \) implies that \( x_1 = x_2 \).
Consider a real-valued injective function $f$. For an available channel set ${\bf c}=\{c_1,c_2, \ldots, c_n\}$, let $f({\bf c})=\{f(c_1), f(c_2), \ldots, f(c_n)\}$. Then, the channel selection function that selects the channel $c$ with the smallest value of $f(c)$ among the set $f({\bf c})$, i.e.,
\beq{small1111}
\phi({\bf c})=\mbox{argmin}_{c \in {\bf c}}f({\bf c}),
\eeq
is also a consistent channel selection function.
In light of \req{small1111}, one can see that the LSH algorithms in \cite{LSH} are based on consistent channel selection functions (that uses a certain hash function $f$).
\eex

In the following lemma, we show that the consistency property is preserved through channel relabeling (with a permutation $\pi$). 

\blem{perm}
Let $\pi$ be a permutation of $\{1,2,\ldots, N\}$ and $\pi^{-1}$ be its inverse permutation. For an available channel set ${\bf c}=\{c_1,c_2, \ldots, c_n\}$, let $\pi({\bf c})=\{\pi(c_1), \pi(c_2), \ldots, \pi(c_n)\}$. If $\phi$ is a consistent channel selection function, then
$\tilde \phi=\pi^{-1} \circ \phi \circ \pi$ (with $\tilde \phi({\bf c})=\pi^{-1}(\phi(\pi({\bf c})))$) is also a consistent channel selection function.
\elem

\bproof
We need to show the condition in \req{consist1111} holds.
Suppose that ${\bf c}^\prime \subset {\bf c}^{\prime\prime}$, and  $\tilde \phi({\bf c}^{\prime\prime})=c$ for some $c$ in ${\bf c}^\prime$.
Thus,
$$\tilde \phi({\bf c}^{\prime\prime})=\pi^{-1}(\phi(\pi({\bf c}^{\prime\prime})))=c.$$
This implies that
\beq{consist2222}
\phi(\pi({\bf c}^{\prime\prime}))=\pi(c).
\eeq
Since $c$ is  in ${\bf c}^\prime$, we know that $\pi(c)$ is in $\pi({\bf c}^\prime)$.
Also, as ${\bf c}^\prime\subset {\bf c}^{\prime\prime}$, $\pi({\bf c}^\prime) \subset \pi({\bf c}^{\prime\prime})$.
Since $\phi$ is a consistent channel selection function,
 \beq{consist3333}
\phi(\pi({\bf c}^\prime))=\phi(\pi({\bf c}^{\prime\prime}))=\pi(c).
\eeq
This then leads to
$$\tilde \phi({\bf c}^\prime)=\pi^{-1}(\phi(\pi({\bf c}^\prime)))=c.$$
Thus, the condition in \req{consist1111} holds.
\eproof

In the following theorem, we show that there are $N!$ distinct consistent channel selection functions.

\bthe{numberf}
There are $N!$ distinct consistent channel selection functions when there are $N$ distinct channels.
\ethe

\bproof
We prove this by induction.
Clearly, for $N=1$, there is exactly one trivial channel selection function, i.e., $\phi(\{1\})=1$.

Assume the induction hypothesis holds, that there are $N!$ distinct consistent channel selection functions when there are $N$ distinct channels.
When there are $N+1$ distinct channels, there are $N+1$ ways to specify $\phi(\{1,2, \ldots, N+1\})$.
Without loss of generality, assume that $\phi(\{1,2, \ldots, N+1\})=1$. Then it follows from the consistency property in \req{consist1111}
that $\phi({\bf c})=1$ for any subset $\bf c$ of $\{1,2, \ldots, N+1\}$ with $1 \in {\bf c}$. It remains to specify $\phi({\bf c})$ for ${\bf c}$ that does not contain 1, and this reduces to the setting with $N$ distinct channels. From the induction hypothesis, there are $N!$
distinct consistent channel selection functions to specify $\phi({\bf c})$ for ${\bf c}$ that does not contain 1.
As such, there are $(N+1)!$ distinct consistent channel selection functions when there are $N+1$ distinct channels.
\eproof

According to \rlem{perm}, we know that these  $N!$ consistent channel selections must all correspond to $\phi^{\min}$ through channel relabeling.
To establish the explicit connection,
we note that the proof of \rthe{numberf} also provides a method to specify a consistent channel selection function with a permutation $\sigma=(\sigma(1),\sigma(2), \ldots, \sigma(N))$.
Define the consistent channel selection function $\phi$ that satisfies
\beq{numberf1111}\phi(\{1,2,\ldots, N\} \backslash  \cup_{j=1}^{i-1} \{ \sigma(j)\})=\sigma(i),
\eeq
where  $i=1,2,\ldots, N$.
In particular, the consistent channel selection function $\phi^{\min}$ (which always selects the channel with the smallest index from the available channel set) corresponds to the permutation $\sigma$ with $\sigma(i)=i$, $i=1,2, \ldots, N$.
By taking the inverse permutation $\sigma^{-1}$ on both sides of \req{numberf1111}, we have
\beq{numberf2222}
\sigma^{-1}\Big (\phi(\{1,2,\ldots, N\} \backslash  \cup_{j=1}^{i-1} \{ \sigma(j)\})\Big)=i .
\eeq
This shows that $\phi^{\min}= \sigma^{-1} \circ \phi \circ \sigma$.
Thus, all the $N!$ consistent channel selection functions can be transformed into $\phi^{\min}$
through channel relabeling with $\pi=\sigma^{-1}$. This is stated in the following corollary.

\bcor{represent}
Every consistent channel selection function of the $N$ channels $\{1,2, \ldots, N\}$ can be represented by
$\pi^{-1} \circ \phi^{\min} \circ \pi$, where $\pi$ is a permutation of $\{1,2, \ldots, N\}$.
\ecor

For instance, $\phi^{\max}$ (that selects  the channel with the largest index from the available channel set) can be represented by $\pi^{-1} \circ \phi^{\min} \circ \pi$ with $\pi(i)=N+1-i$, $i=1,2, \ldots, N$.

\bsubsec{Consistent channel hopping algorithms}{ccha}

In this section, we show how consistent channel selection functions can be used for the multichannel rendezvous problem.

\bdefin{ccha}
A channel hopping algorithm  is said to be {\em consistent} if it uses a sequence of consistent channel selection functions
$\{\phi_1, \phi_2, \ldots\}$ to generate its channel hopping sequence, where the channel selected at time $t$ from an available channel set ${\bf c}$ is $\phi_t({\bf c})$.
\edefin

In the following theorem, we provide a necessary and sufficient characterization of the set of rendezvous time slots.

\bthe{rendezvous}
Consider the MRP in the synchronous setting.
Assume that  the two users use a consistent channel hopping algorithm $\{\phi_1, \phi_2, \ldots\}$ to generate their channel hopping sequences.
Let ${\bf c}_1$ (resp. ${\bf c}_2$) be the available channel set of user 1 (resp. 2).
Consider a {\em fictitious} user, referred to as user 0, with the available channel set ${\bf c}_1 \cup {\bf c}_2$.
Let  $c_0(t)=\phi_t({\bf c}_1 \cup {\bf c}_2)$ (resp. $c_1(t)=\phi_t({\bf c}_1)$ and $c_2(t)=\phi_t({\bf c}_2)$) be the channel selected by user 0 (resp. user 1 and user 2) at time $t$.
Denote by ${\cal T}_i$ the set of time slots that user 0 selects channel $i$, and by
${\cal T}=
\cup_{i \in {\bf c}_1 \cap {\bf c}_2}{\cal T}_i$  the set of time slots that user 0 selects a common channel.
Then the two users rendezvous at time $t$ if and only if $t \in {\cal T}$.
\ethe

\bproof
We first prove the {\em if} direction.
Suppose that channel $i$ is a  common channel, i.e., $i \in {\bf c}_1 \cap {\bf c}_2$, and that channel $i$ is selected
at time $t$ by user 0. Then it follows
from the definition of a consistent channel selection function that channel $i$ is also selected by user 1 and user 2 at time $t$. Thus, the two users
will rendezvous at channel $i$ at time $t$.

Now we prove the {\em only if} direction by contradiction. Suppose that channel $i$ is a  common channel and that
the two users rendezvous at channel $i$ at some time $t \not \in {\cal T}_i$. Assume that user 0 selects channel $j \ne i$ at time $t$.
Since $j \in {\bf c}_1 \cup {\bf c}_2$, we consider the following three cases. First, if $j \in {\bf c}_1 \cap {\bf c}_2$,
then it follows
from the definition of a consistent channel selection function that channel $j$ is also selected by user 1 and user 2 at time $t$.
This contradicts the assumption that the two users rendezvous at channel $i$ at time $t$. Second, if $j \in {\bf c}_1$ and $j \not \in {\bf c}_2$,
then it follows
from the definition of a consistent channel selection function that channel $j$ is selected by user 1 at time $t$. Since $j \ne i$, this contradicts the assumption that the two users rendezvous at channel $i$ at time $t$. Finally, if $j \not \in {\bf c}_1$ and $j \in {\bf c}_2$, then
it follows
from the definition of a consistent channel selection function that channel $j$ is selected by user 2 at time $t$. Once again,
 this contradicts the assumption that the two users rendezvous at channel $i$ at time $t$.
\eproof

\bdefin{pi}
Given a sequence of permutations ${\bf \Pi}=\{\pi_1,\pi_2, \ldots\}$,
 the channel hopping algorithm $\{\phi_1, \phi_2, \ldots\}$ with $\phi_t = \pi_t^{-1} \circ \phi^{\min} \circ \pi_t$, $t =1,2, \ldots$, is called
the ${\bf \Pi}$-algorithm.
\edefin

In view of \rcor{represent}, we have the following corollary.

\bcor{pi}
Every consistent channel hopping algorithm can be represented by a ${\bf \Pi}$-algorithm.
\ecor

Now we show how to find consistent channel hopping algorithms with tight MTTR bounds and good ETTR.

\bdefin{onecylce}
For a permutation $\pi$, define \( \pi^{i+1} \) as the composition of \( \pi^i \) with \( \pi \) (i.e., applying \( \pi \) repeatedly for $i+1$ times).
A one-cycle permutation is a permutation that consists of a single cycle of length
$N$. In such a permutation, the elements are cyclically permuted, and $\pi^N$
  is the identity permutation, i.e., for any $i$ in \( \{1, 2, \ldots, N\} \), $(\pi(i), \pi^2(i), \ldots, \pi^N(i))$ is a permutation of \( \{1, 2, \ldots, N\} \) and $\pi^N(i)=i$.
\edefin

\bex{onecycle}{}
It is easy to see that  the cyclic shift permutation  $(2,3,\ldots, N,1)$, i.e., $\pi(i)=(i \;\mbox{mod} \;N)+1$ for $i=1,2,\ldots, N$, is a one-cycle permutation. Such a permutation is also known as a rotation and we denote it  as $\pi^{\rm rot}$.
\eex

In the following theorem, we derive an MTTR bound by using a ${\bf \Pi}$-algorithm generated by a one-cycle permutation.

\bthe{MTTR}
Consider the MRP in the synchronous setting.
Assume that  the two users generate their channel hopping sequences by using the ${\bf \Pi}$-algorithm with ${\bf \Pi}=\{\pi,\pi^2, \ldots\}$, where $\pi$ is a one-cycle permutation.
Then the two users rendezvous on every common channel and
the MTTR is bounded by $N-n_{1,2}+1$.
\ethe

The proof of \rthe{MTTR} relies on the following lemma.

\blem{once}
Consider the ${\bf \Pi}$-algorithm with ${\bf \Pi}=\{\pi,\pi^2, \ldots\}$, where $\pi$ is a one-cycle permutation.
Then the channel hopping sequence of such an algorithm is periodic with period $N$. Moreover,
for any available channel set ${\bf c}$, every channel appears at least once in every $N$ consecutive time slots of the channel hopping sequence.
\elem

\bproof
Since $\pi$ is a one-cycle permutation, we know that $\pi^N$ is the identity permutation, i.e., $\pi^N(i)=i$ for all $i=1,2, \ldots, N$.
The fact that the channel hopping sequence is periodic with period $N$ follows from $\pi^{N+1}=\pi$.

Now we show that every channel appears at least once in a period of $N$ time slots.
Denote by $\Omega=\{1,2, \ldots, N\}$.
We first show  that
\beq{once1111} \{(\pi^{-1} \circ \phi^{\min} \circ \pi)(\Omega), \ldots, ((\pi^N)^{-1} \circ \phi^{\min} \circ \pi^N)(\Omega)\} =\Omega.
\eeq
As $\pi$ is a one-cycle permutation, we have $(\pi^i)^{-1} =\pi^{N-i}$ for all $i=1,2, \ldots, N$.
Thus,
$$\Big((\pi^i)^{-1} \circ \phi^{\min} \circ \pi^i\Big )(\Omega) =\pi^{N-i}(\phi^{\min}(\Omega)).$$
Since $\phi^{\min}$ always selects the smallest indexed channel, we have $\phi^{\min}(\Omega)=1$.
As $\pi$ is a one-cycle permutation, we have
\bearn
&&\{(\pi^{-1} \circ \phi^{\min} \circ \pi)(\Omega),  \ldots, ((\pi^N)^{-1} \circ \phi^{\min} \circ \pi^N)(\Omega)\}\\
&&=\{\pi^{N-1}(1), \pi^{N-2}(1), \ldots, \pi(1), 1\}=\Omega.
\eearn

Consider a channel $c$ in ${\bf c}$. From \req{once1111}, there exists a $1 \le t \le N$ such that
$$((\pi^t)^{-1} \circ \phi^{\min} \circ \pi^t)(\Omega)=c.$$
Since ${\bf c}$ is a subset of $\Omega$, we know from the consistent property (i.e., the condition in \req{consist1111}) that
$$((\pi^t)^{-1} \circ \phi^{\min} \circ \pi^t)({\bf c})=c.$$
This shows that every channel of ${\bf c}$ appears at least once in the first $N$ time slots.
\eproof

\bproof(\rthe{MTTR})
From \rthe{rendezvous}, the set of rendezvous time slots is the set of time slots that user 0 (the fictitious user) selects a common channel $c \in {\bf c}_1 \cap {\bf c}_2$. The available channel set for user 0 is ${\bf c}_1 \cup {\bf c}_2$.
It then follows from \rlem{once} that every common channel appears at least once in the first $N$ time slots.
Since there are $n_{1,2}$ common channels, the MTTR is bounded by $N-n_{1,2}+1$.
\eproof

In the following theorem, we show that the ETTR is the inverse of the Jaccard index for a ${\bf \Pi}$-algorithm generated by random permutations.

\bthe{fair}
Consider the MRP in the synchronous setting. Let ${\bf c}_1$ (resp. ${\bf c}_2$) be the available channel set of user 1 (resp. user 2).
Assume that  the two users use the ${\bf \Pi}$-algorithm  to generate their channel hopping sequences, where $\{\pi_t$, $t=1,2, \ldots\}$ are random permutations (that are independently and uniformly selected from the $N!$ permutations).
Then the ETTR is $1/J$,
where $J=\frac{n_{1,2}}{n_{1} + n_{2} - n_{1,2}}$ is the Jaccard index between ${\bf c}_1$ and ${\bf c}_2$.
\ethe

We note that the sequence of  independently and uniformly selected random permutations $\{\pi_t, t =1, 2, \ldots\}$ has to be shared by the two users.
This can be achieved using pseudo-random permutations with a common seed.

The proof of \rthe{fair} relies on the following lemma.

\blem{fairgen}
Suppose $\pi$ is a random permutation. Then for any channel selection function $\phi$ and any available channel set $\bf c$,
\beq{fair0011}
\pr (\phi(\pi({\bf c}))=\pi(c))= \frac{1}{|\bf c|}.
\eeq
\elem

\bproof
Since $\pi$ is a random permutation,
\beq{fair0022}
\pr (\phi(\pi({\bf c}))=\pi(c)) =\frac{1}{N!} \sum_{\sigma} {\bf 1}_{\{\phi(\sigma({\bf c}))=\sigma(c)\}},
\eeq
where the sum over $\sigma$ is over all the $N!$ permutations.
Thus, it suffices to show that for any $c_1 \ne c_2$ in ${\bf c}$,
\beq{fair0033}
\sum_{\sigma} {\bf 1}_{\{\phi(\sigma({\bf c}))=\sigma(c_1)\}}=\sum_{\sigma} {\bf 1}_{\{\phi(\sigma({\bf c}))=\sigma(c_2)\}}.
\eeq
Suppose for some permutation $\sigma$, $\phi(\sigma({\bf c}))=\sigma(c_1)$. We will show that there is a corresponding permutation $\sigma^\prime$ such that $\phi(\sigma^\prime({\bf c}))=\sigma^\prime(c_2)$.
Specifically, we consider the permutation $\sigma^\prime$ that swaps \( \sigma(c_1) \) and \( \sigma(c_2) \) and fixes all other elements, i.e., $\sigma^\prime(c_1)=\sigma(c_2)$, $\sigma^\prime(c_2)=\sigma(c_1)$, and $\sigma^\prime(i)=\sigma(i)$ for $i \ne c_1, c_2$.
Clearly, for $c_1 \ne c_2$ in ${\bf c}$, we have $\sigma^\prime({\bf c})=\sigma({\bf c})$.
As such,
$$\phi(\sigma^\prime({\bf c}))=\phi(\sigma({\bf c}))=\sigma (c_1)=\sigma^\prime(c_2).$$
Thus, \req{fair0033} holds.
\eproof

\bproof(\rthe{fair})
From \rthe{rendezvous}, the set of rendezvous time slots is the set of time slots that user 0 selects a common channel $c \in {\bf c}_1 \cap {\bf c}_2$. The available channel set for user 0 is ${\bf c}_1 \cup {\bf c}_2$.
As a result of  \rlem{fairgen}, the probability  that the two users rendezvous at time $t$ is
$$\pr \Big (\phi_t({\bf c}_1 \cup {\bf c}_2) \in {\bf c}_1 \cap {\bf c}_2 \Big )= \frac{|{\bf c}_1 \cap {\bf c}_2|}{|{\bf c}_1 \cup {\bf c}_2|}=J.$$ Since the channel hopping sequence of user 0 is an i.i.d. sequence, the time-to-rendezvous is geometrically distributed with mean $1/J$. Thus, the ETTR is $1/J$.
\eproof

\bsubsec{Optimality of consistent channel hopping algorithms}{optccha}

In this section, we show that consistent channel hopping algorithms maximize the  rendezvous probability at any time $t$.
Consider a general channel hopping algorithm $\{\phi_1, \phi_2, \ldots \}$, where $\phi_t$'s are not necessarily consistent channel selection functions. Define the rendezvous probability at time $t$ as the fraction of times that  the two users rendezvous at time $t$ over all the $N!$ ways of channel relabeling (through a permutation $\sigma$), i.e.,

\beq{fair0044}
\frac{1}{N!}\sum_{\sigma} {\bf 1}_{\{\phi_t(\sigma({\bf c_1}))=\phi_t(\sigma({\bf c_2}))\}}.
\eeq

\bthe{ETTRgen}
For any channel hopping algorithm $\{\phi_1, \phi_2, \ldots \}$,
the rendezvous probability at time $t$ cannot be larger than $J$, and it is $J$ if
the channel hopping algorithm is {\em consistent}.
\ethe


\bproof
Note that \req{fair0044} can be rewritten as follows:
\beq{fair0055}
\pr \Big (\phi_t(\pi({\bf c_1}))=\phi_t(\pi({\bf c_2}))\Big),
\eeq
where $\pi$ is a random permutation.
From the law of total probability, we then have
\bear{fair0066}
&&\pr \Big (\phi_t(\pi({\bf c_1}))=\phi_t(\pi({\bf c_2}))\Big) \nonumber\\
&&=\sum_{c \in {\bf c}_1 \cup {\bf c}_2} \pr \Big (\phi_t(\pi({\bf c_1}))=\phi_t(\pi({\bf c_2}))=\pi(c) \nonumber\\ &&\quad\quad\quad\quad \Big | \phi_t(\pi({\bf c_1} \cup {\bf c}_2))=\pi(c)  \Big ) \nonumber\\
&& \quad\quad\quad\quad  \pr \Big (\phi_t(\pi({\bf c_1} \cup {\bf c}_2))=\pi(c) \Big) .
\eear
Since
\bearn
&&\pr \Big (\phi_t(\pi({\bf c_1}))=\phi_t(\pi({\bf c_2}))=\pi(c) \nonumber\\ &&\quad\quad\quad\quad \Big | \phi_t(\pi({\bf c_1} \cup {\bf c}_2))=\pi(c)  \Big )=0,
\eearn
for $c \not \in {\bf c}_1 \cap {\bf c}_2$,
we have
\bear{fair0066a}
&&\pr \Big (\phi_t(\pi({\bf c_1}))=\phi_t(\pi({\bf c_2}))\Big) \nonumber\\
&&=\sum_{c \in {\bf c}_1 \cap {\bf c}_2} \pr \Big (\phi_t(\pi({\bf c_1}))=\phi_t(\pi({\bf c_2}))=\pi(c) \nonumber\\ &&\quad\quad\quad\quad \Big | \phi_t(\pi({\bf c_1} \cup {\bf c}_2))=\pi(c)  \Big ) \nonumber\\
&& \quad\quad\quad\quad \pr \Big (\phi_t(\pi({\bf c_1} \cup {\bf c}_2))=\pi(c) \Big) .
\eear
From \rlem{fairgen}, we have
\beq{fair0077}
\pr \Big (\phi_t(\pi({\bf c_1} \cup {\bf c}_2))=\pi(c) \Big)=\frac{1}{|{\bf c_1} \cup {\bf c}_2|}.
\eeq
On the other hand, as the conditional probability cannot be greater than 1,
\bear{fair0088}
&&\pr \Big (\phi_t(\pi({\bf c_1}))=\phi_t(\pi({\bf c_2}))=\pi(c) \nonumber\\ &&\quad\quad\quad\quad \Big | \phi_t(\pi({\bf c_1} \cup {\bf c}_2))=\pi(c)  \Big )\le 1.
\eear
This shows that
\beq{fair0099}
\pr \Big (\phi_t(\pi({\bf c_1}))=\phi_t(\pi({\bf c_2}))\Big) \le \frac{|{\bf c_1} \cap {\bf c}_2|}{|{\bf c_1} \cup {\bf c}_2|}=J.
\eeq
 Thus, the rendezvous probability at time $t$ cannot be larger than
$J$ for any channel selection function $\phi_t$. In view of \req{fair0088}, one can achieve the rendezvous probability $J$ if  $\phi_t$ is a consistent channel selection function (according to the proof of \rthe{fair}).
\eproof

\bsec{Implementations of ${\bf \Pi}$-algorithms}{exp}

In the previous section, we showed that a consistent channel hopping algorithm can be characterized by a sequence of permutations ${\bf \Pi} = \{\pi_1, \pi_2, \ldots\}$. In this section, we discuss design principles for constructing a good sequence of permutations.

\begin{description}
\item[(i)] \textbf{MTTR:} In \rthe{MTTR}, we show that a tight MTTR bound can be achieved if the sequence of permutations is generated using a one-cycle permutation.
\item[(ii)] \textbf{ETTR:} In \rthe{fair}, we show that the ETTR is $1/J$ when the sequence of permutations is generated by random permutations.
\item[(iii)] \textbf{Computational Complexity:} Generating random permutations is costly when the total number of channels $N$ is very large. If the number of available channels $n$ of a user is much smaller than $N$, it is desirable to generate the channel hopping sequence in a way that depends only on $n$.
\end{description}

Based on these, we seek one-cycle permutations that are easy to implement and can generate ``random-like'' permutations, ensuring both a good MTTR bound and a good ETTR.

\bsubsec{The locality-sensitive hashing based algorithms}{LSH}

In this section, we show that LSH2 in \cite{LSH} is an implementation of a ${\bf \Pi}$-algorithm.
LSH2 uses two permutations $\pi_1$ and $\pi_2$ and generates the channel hopping sequence $\{c(t), t=1,2, \ldots, N\}$
for an available channel set ${\bf c}=\{c_1, c_2, \ldots, c_n\}$ by selecting
$c(t)=c_{i^*}$, where  $i^*$ is the index of the available channel such that the difference between $\pi_1(c_i^*)$ and $\pi_2(t)$ is the smallest in the modulo manner among all the $n$ available channels,
i.e.,
 $$i^*={\rm argmin}_{1 \le i \le n}((\pi_1(c_i)-\pi_2(t))\;\mod\;N).$$
Note that $\pi_1$ is for relabeling of the $N$ channels and $\pi_2$ is for relabeling of the $N$ time slots.
These two relabeling permutations constitute a composite relabeling of $N$ channels for each time slot $t$.
Such a composite relabeling can be represented by
\beq{LSH2equ1111}
\pi_t=((\pi^{\rm rot})^{-1})^{\pi_2(t)} \circ \pi_1,
\eeq
where $\pi^{\rm rot}$ is the cyclic shift permutation in \rex{onecycle}.
Thus, LSH2 is equivalent to the ${\bf \Pi}$-algorithm with
\beq{LSH2equ2222}
\phi_t= (\pi_t)^{-1} \circ \phi^{\min} \circ \pi_t.
\eeq
If we simply choose $\pi_1$ and $\pi_2$ to be the identity permutation,  then
the LSH2 algorithm is the ${\bf \Pi}$-algorithm generated by the
 one-cycle permutation $(\pi^{\rm rot})^{-1}$. The objective of  $\pi_2$ is to shuffle the
 $N$ permutations $\{((\pi^{\rm rot})^{-1} )^t, t=1,2, \ldots, N\}$ so that they appear to be a
 sequence of random permutations.

\bsubsec{The modulo algorithm}{algorithm}

The LSH2 algorithm in \cite{LSH} require implementing pseudo-random permutations $\pi_1$ and $\pi_2$.
When $N$ is large, this implementation incurs a high computational cost.
It is well-known that permutations can be easily computed using the modulo operation with respect to a prime number.
To avoid generating pseudo-random permutations,
 we propose using the following one-cycle permutation $\pi$ with
 \beq{onecylce1111}
 \pi(i)=(\gen \cdot i \;\mbox{mod}\;P)
 \eeq
  when $P=N+1$ is a prime and $\gen$ is a generator of the multiplicative group $\{1,2, \ldots, P-1\}$.
  Such a ${\bf \Pi}$-algorithm is called the modulo algorithm as it uses the modulo operation to generate permutations.
    Recall that a generator $g$ generates all the elements in $\{1,2, \ldots, P-1\}$, i.e., $\{g^1, g^2, \ldots, g^{P-1}\}=\{1,2, \ldots, P-1\}$ and there is always a generator when $P$ is a prime. For instance, when $P=257$,  the numbers 3, 126, and 254 are all generators.
 In case that $N+1$ is not a prime, we can always add fictitious channels to the total channels so that $N+1$
   is a prime.
 Note that adding fictitious channels increases the period of the one-cycle permutation and might lead to degradation of performance for MTTR.

For such a one-cycle permutation,
   we have
   \bear{onecylce2222}
   \pi^t(i)&=&(\gen^t \cdot i \;\mbox{mod}\;P), \\
    (\pi^t)^{-1}(i)&=&(\gen^{P-1-t} \cdot i \;\mbox{mod}\;P).
    \eear

For the modulo algorithm,
the channel selected at time $t$ for an available channel set ${\bf c}$
is
\bear{onecycle4444}
&&\Big ((\pi^t)^{-1} \circ \phi^{\min}  \circ \pi^t \Big)({\bf c})= (\pi^t)^{-1}( \phi^{\min}(\pi^t ({\bf c})))\nonumber\\
&&={\rm argmin}_{c \in {\bf c}}(\gen^t \cdot c \;\mbox{mod}\;P).
\eear
This can be recursively computed by keeping each available channel $c$ a virtual clock $v_t(c)$
with $v_0(c)=c$ and
\beq{onecycle5555}
v_{t+1}(c)=(\gen \cdot v_t(c)  \;\mbox{mod}\;P).
\eeq
Note that \req{onecycle5555} only needs to be computed for each available channel. Thus, the computational complexity is only $O(n)$ in each time slot, where $n$ is the number of available channels.

We note that the performance of the modulo algorithm is influenced by the choice of the generator. If the generator is too small, it may fail to select channels evenly, resulting in a non-uniform probability distribution.
For example, consider \(P = 257\) and \(g = 3\), with the available channel set \(\{1, 3, 9, 27\}\) over one period. In this case, channels 3, 9, and 27 will each be selected exactly once, while channel 1 will be selected 253 times. This violates the design principle that the channel selection sequence should exhibit random-like behavior.

\bsec{Multichannel rendezvous with multiple users}{musers}

In this section, we consider the MRP with multiple users.
Suppose that there are $K$ users with the available channel sets ${\bf c}_k$, $k=1,2,\ldots, K$.
As in \req{avail1111}, we assume that there is at least one channel that is available to the $K$ users, i.e.,
\beq{avail1111K}
\cap_{k=1}^K {\bf c}_k \ne \varnothing.
\eeq
The TTR is defined as the number of time slots needed for $K$ users to hop to a common channel at the same time.

\bsubsec{Direct extensions}{direct}

In this section, we assume that each user follows the ${\bf \Pi}$-algorithm to generate its channel hopping sequence. The result in \rthe{rendezvous}, which characterizes the rendezvous time slots for two users, can be readily extended to the case with $K$ users as follows:

\bcor{rendezvousK}
Consider the MRP in the synchronous setting with $K$ users.
Assume that  the $K$ users use a consistent channel hopping algorithm $\{\phi_1, \phi_2, \ldots\}$ to generate their channel hopping sequences.
Consider a {\em fictitious} user, referred to as user 0, with the available channel set $\cup_{k=1}^K {\bf c}_k$.
Denote by ${\cal T}_i$ the set of time slots that user 0 selects channel $i$, and by
${\cal T}=\cup_{i \in \cap_{k=1}^K {\bf c}_k}{\cal T}_i$  the set of time slots that user 0 selects a channel in $\cap_{k=1}^K {\bf c}_k$.
Then the $K$ users rendezvous at time $t$ if and only if $t \in {\cal T}$.
\ecor

The MTTR bound in \rthe{MTTR} is still valid for a ${\bf \Pi}$-algorithm generated by a one-cycle permutation.

\bcor{MTTRK}
Consider the MRP in the synchronous setting with $K$ users.
Assume that  the $K$ users generate their channel hopping sequences by using the ${\bf \Pi}$-algorithm with ${\bf \Pi}=\{\pi,\pi^2, \ldots\}$, where $\pi$ is a one-cycle permutation.
Then the $K$ users rendezvous on every common channel and
the MTTR is bounded by $N-|\cap_{k=1}^K {\bf c}_k|+1$.
\ecor

Define the (generalized) Jaccard index
$J$ as $|\cap_{k=1}^K {\bf c}_k|/|\cup_{k=1}^K {\bf c}_k|$.
Similar to the result in \rthe{fair}, the ETTR is the inverse of the Jaccard index for a ${\bf \Pi}$-algorithm generated by random permutations.

\bcor{fairK}
Consider the MRP in the synchronous setting.
Assume that  the $K$ users use the ${\bf \Pi}$-algorithm  to generate their channel hopping sequences, where $\{\pi_t$, $t=1,2, \ldots\}$ are random permutations (that are independently and uniformly selected from the $N!$ permutations).
Then the ETTR is $1/J$,
where
\beq{fairK2222}
J=\frac{|\cap_{k=1}^K {\bf c}_k|}{|\cup_{k=1}^K {\bf c}_k|}.
\eeq
\ecor

\bsubsec{The stick-together strategy}{stick}

In the direct extension of the MRP with $K$ users in \rsec{direct},  users' channel hopping sequences remain unchanged even when a subset of users rendezvous at a certain time $t$. In fact, when a subset of users rendezvous,  they not only discover the existence of each other but also can exchange information \cite{liu2012jump}.
In particular, the ``stick-together'' strategy in \cite{ToN2017} can take advantage of this to speed  up the rendezvous process. To be specific,
suppose that a subset $S$ of the $K$ users rendezvous at time $t$. Then these users can exchange the information of
their available channel sets and adopt the new available channel set $\cap_{k \in S}{\bf c}_k$ for their channel hopping sequences after time $t$. As such, after time $t$, the MRP with $K$ users is reduced to the usual MRP with $K-|S|+1$ users.
In the following theorem, we prove that the TTR with the stick-together strategy is not larger than that  without changing channel hopping
sequences.

\bthe{stick}
Consider the MRP in the synchronous setting with $K$ users.
Assume that  the $K$ users use a consistent channel hopping algorithm $\{\phi_1, \phi_2, \ldots\}$ to generate their channel hopping sequences.
Let $T^{\rm uch}$ be the TTR for the strategy without changing the channel hopping sequences of the  $K$ users (in \rsec{direct}),
and
$T^{\rm st}$ be the TTR for the stick-together strategy that changes the channel hopping sequences when a subset of users rendezvous (as described in this section). Then $T^{\rm st} \le T^{\rm uch}$.
\ethe

\bproof

From \rcor{rendezvousK}, the rendezvous time slots for the strategy without changing the channel hopping sequences are
the time slots that the fictitious user, i.e., user 0, with  the available channel set $\cup_{k=1}^K {\bf c}_k$,
  selects a channel in $\cap_{k=1}^K {\bf c}_k$.
  As in \rcor{rendezvousK}, we let ${\cal T}_i$ be the set of time slots that user 0 selects channel $i$, and
${\cal T}=\cup_{i \in \cap_{k=1}^K {\bf c}_k}{\cal T}_i$  be the set of time slots that user 0 selects a channel in $\cap_{k=1}^K {\bf c}_k$.
It suffices to show that every rendezvous time slot for the strategy without changing the channel hopping sequences is still a
rendezvous time slot for the stick-together strategy. To see this, note that every channel in $\cap_{k=1}^K {\bf c}_k$ is still in the new available channel set $\cap_{k \in S}{\bf c}_k$ when a subset $S$ of users rendezvous at time $t$.
It then follows from the consistency property in \rdef{ccsf} that the channel selected by the
stick-together strategy remains unchanged for every time slot in ${\cal T}$. Thus, every rendezvous time slot for the strategy without changing the channel hopping sequences is also a rendezvous time slot for the stick-together strategy.
\eproof

\bsubsec{Exact analysis for the ETTR of the stick-together strategy when $K=3$}{ETTR3}

In this section, we derive the ETTR for three users, each of whom adopts the ${\bf \Pi}$-algorithm generated by a sequence of random permutations.
The ETTR of the strategy without changing the channel hopping sequences follows directly from \rcor{fairK}, i.e.,
\beq{fairK1111}
\ex[ T^{\rm uch}] =1/J,
\eeq
where $J$ is the (generalized) Jaccard index in \req{fairK2222}. 

Now we derive the ETTR for the stick-together strategy.
 Note that for $K=2$, $T^{\rm uch}=T^{\rm st}$ and
 the ETTR of the stick-together strategy can be computed by using \rthe{fair}.

Now we use a Markov chain analysis to compute $\ex[ T^{\rm st}]$ for $K=3$.
A state $(K, \{{\bf c}_1, \ldots, {\bf c}_K\})$ in the Markov chain denotes the MRP with $K$ users and the corresponding available channel sets ${\bf c}_1, \ldots , {\bf c}_K$. Let $T_{(K, \{{\bf c}_1, \ldots, {\bf c}_K\})}$ be the TTR starting from the state $(K, \{{\bf c}_1, \ldots, {\bf c}_K\})$ under the stick-together strategy.
For the MRP with three users, the initial state is $(3, \{{\bf c}_1, {\bf c}_2, {\bf c}_3\})$.
Let $n_k$ be the number of channels in ${\bf c}_k$, $k=1,2,$ and 3, $n_{1,2}$ (resp. $n_{1,3}$ and $n_{2,3}$)
be the number of channels in ${\bf c}_1 \cap {\bf c}_2$ (resp. ${\bf c}_1 \cap {\bf c}_3$ and
${\bf c}_2 \cap {\bf c}_3$), and $n_{1,2,3}$ be the number of channels in ${\bf c}_1 \cap {\bf c}_2 \cap {\bf c}_3$.
Also, we let  $n^{\rm union}$ be the number of channels in ${\bf c}_1 \cup {\bf c}_2 \cup {\bf c}_3$. Note from the inclusion-exclusion principle that
\beq{fairK2233}
n^{\rm union}=n_1+n_2+n_3-n_{1,2}-n_{1,3}-n_{2,3}+n_{1,2,3}.
\eeq
Also, we denote ${\bar {\bf c}}$ as the complement of ${\bf c}$, i.e., $\{1,2, \ldots, N\}\backslash {\bf c}$.

As the ${\bf \Pi}$-algorithms of the three users are generated by random permutations, we can condition on the events at time 1.   There are five possible events at time 1: (i) user 1 and user 2 rendezvous at time 1, but user 3 does not (denoted by the event $E_{1,2,\bar 3}$),  (ii) user 1 and user 3 rendezvous at time 1, but user 2 does not (denoted by the event $E_{1,\bar 2,3}$),  (iii) user 2 and user 3 rendezvous at time 1, but user 1 does not (denoted by the event $E_{\bar 1,2,3}$),
(iv) the three users rendezvous at time 1 (denoted by the event $E_{1,2,3}$), and (v) none of the above (denoted by the event $E_0$).

When the event $E_{1,2,\bar 3}$ (resp. $E_{1,\bar 2,3}$ and $E_{\bar 1, 2,3}$) and happens, the Markov chains transitions into the new state $(2, \{{\bf c}_1 \cap {\bf c}_2, {\bf c}_3\})$ 
(resp. $(2, \{{\bf c}_1 \cap {\bf c}_3, {\bf c}_2\})$ and $(2, \{{\bf c}_2 \cap {\bf c}_3, {\bf c}_1\})$) under the stick-together strategy.
We have from \rthe{fair} that
\beq{fairK4444b}
\ex[T_{(2, \{{\bf c}_1 \cap {\bf c}_2, {\bf c}_3\})}]
=\frac{n_{1,2}+n_3 -n_{1,2,3}}{n_{1,2,3}},
\eeq
\bear{fairK5555b}
&& \ex[T_{(2, \{{\bf c}_1 \cap {\bf c}_3, {\bf c}_2\})}]=\frac{n_{1,3}+n_2 -n_{1,2,3}}{n_{1,2,3}},
\eear
and 
\bear{fairK6666b}
&& \ex[T_{(2, \{{\bf c}_2 \cap {\bf c}_3, {\bf c}_1\})}]=\frac{n_{2,3}+n_1 -n_{1,2,3}}{n_{1,2,3}}.
\eear
When the event $E_{1, 2,3}$ happens, all the three users rendezvous.
When the event $E_0$ happens, the Markov chain remains in the same state.

Conditioning on these five events at time 1, we derive the following recursive equation:
\bear{fairK8888}
&&\ex[T_{(3, \{{\bf c}_1, {\bf c}_2, {\bf c}_3\})}]=1+ \pr (E_{1,2,\bar 3})\ex[T_{(2, \{{\bf c}_1 \cap {\bf c}_2, {\bf c}_3\})}] \nonumber\\
&&\quad\quad+\pr (E_{1,\bar 2,3})\ex[T_{(2, \{{\bf c}_1 \cap {\bf c}_3, {\bf c}_2\})}]\nonumber\\
&&\quad\quad+\pr (E_{\bar 1, 2,3})\ex[T_{(2, \{{\bf c}_2 \cap {\bf c}_3, {\bf c}_1\})}]\nonumber\\
&&
\quad\quad+\pr (E_0)\ex[T_{(3, \{{\bf c}_1, {\bf c}_2, {\bf c}_3\})}].
\eear
Solving this equation yields
\bear{fairk9999}
&&\ex[T_{(3, \{{\bf c}_1, {\bf c}_2, {\bf c}_3\})}]\nonumber\\
&&=\frac{1}{\pr (E_{1,2,\bar 3})+\pr (E_{1,\bar 2,3})+\pr (E_{\bar 1, 2,3})+\pr (E_{1, 2,3})} \nonumber\\
&&\quad\quad\Big (1+ \pr (E_{1,2,\bar 3})\ex[T_{(2, \{{\bf c}_1 \cap {\bf c}_2, {\bf c}_3\})}] \nonumber\\
&&\quad\quad+\pr (E_{1,\bar 2,3})\ex[T_{(2, \{{\bf c}_1 \cap {\bf c}_3, {\bf c}_2\})}]\nonumber\\
&&\quad\quad+\pr (E_{\bar 1, 2,3})\ex[T_{(2, \{{\bf c}_2 \cap {\bf c}_3, {\bf c}_1\})}]\Big ).\nonumber\\
\eear

In the following lemma, we derive the probabilities of these five events.

\blem{five}
Suppose that the ${\bf \Pi}$-algorithms of the three users are generated by random permutations. Then
\bear{fairK3366}
&&\pr (E_{1,2,\bar 3})
=\frac{n_{1,2}-n_{1,2,3}}{n^{\rm union}}\nonumber\\
&&\quad+\frac{n_{1,2}}{n_1+n_2-n_{1,2}}\frac{n_3-n_{2,3}-n_{1,3}+n_{1,2,3}}{n^{\rm union}},
\eear
\bear{fairK5555}
&&\pr (E_{1,\bar 2,3})=\frac{n_{1,3}-n_{1,2,3}}{n^{\rm union}}\nonumber\\
&&\quad +\frac{n_{1,3}}{n_1+n_3-n_{1,3}}\frac{n_2-n_{2,3}-n_{1,2}+n_{1,2,3}}{n^{\rm union}},
\eear
\bear{fairK6666}
&&\pr (E_{\bar 1, 2,3})=\frac{n_{2,3}-n_{1,2,3}}{n^{\rm union}} \nonumber\\
&&\quad+\frac{n_{2,3}}{n_2+n_3-n_{2,3}}\frac{n_1-n_{1,3}-n_{2,3}+n_{1,2,3}}{n^{\rm union}}, 
\eear
\bear{fairK7777}
&&\pr (E_{1, 2,3})=\frac{n_{1,2,3}}{n^{\rm union}}, 
\eear
and
\bear{fairK7788}
&&\pr (E_0)\nonumber\\
&&=1-\pr (E_{1,2,\bar 3})-\pr (E_{1,\bar 2,3})-\pr (E_{\bar 1, 2,3})-\pr (E_{1, 2,3}).\nonumber\\
\eear
\elem

\bproof
To compute the probabilities of these five events, let us consider the channel hopping sequence of the fictitious user, i.e., user 0, with
the available channel set ${\bf c}_0={\bf c}_1 \cup {\bf c}_2 \cup {\bf c}_3$.

We first show \req{fairK3366}.
Note from the law of total probability that

\bear{fairK3333}
&&\pr (E_{1,2,\bar 3})=\pr \Big (\phi_1({\bf c}_1)= \phi_1({\bf c}_2) \ne \phi_1({\bf c}_3) \Big)\nonumber\\
&&=\sum_{c \in {\bf c}_0}
\pr \Big (\phi_1({\bf c}_1)= \phi_1({\bf c}_2)\ne  \phi_1({\bf c}_3) \Big |
\phi_1({\bf c}_0)=c\Big ) \nonumber\\
&&\quad\quad\quad\quad \pr (\phi_1({\bf c}_0)=c).
\eear
Recall that $\phi_1=\pi_1^{-1} \circ \phi^{\rm min} \circ \pi_1$ with $\pi_1$ being a random permutation.
From \rlem{fairgen},
we know that
\bear{fairK3344}
&&\pr (\phi_1({\bf c}_0)=c)= \pr (\phi^{\rm min}(\pi_1({\bf c}_0))=\pi_1(c))\nonumber\\
&&=\frac{1}{|{\bf c}_1 \cup {\bf c}_2 \cup {\bf c}_3|}.
\eear
To compute the conditional probability in \req{fairK3333}, we partition ${\bf c}_0$ into
seven sets: ${\bf c}_1 \cap {\bf c}_2 \cap {\bf c}_3$,
${\bar {\bf c}}_1 \cap {\bf c}_2 \cap {\bf c}_3$,
${\bf c}_1 \cap {\bar {\bf c}}_2 \cap {\bf c}_3$,
${\bf c}_1 \cap {\bf c}_2 \cap {\bar {\bf c}}_3$,
${\bar {\bf c}}_1 \cap {\bar {\bf c}}_2 \cap {\bf c}_3$,
${\bar {\bf c}}_1 \cap {\bf c}_2 \cap {\bar {\bf c}}_3$, and
${\bf c}_1 \cap {\bar {\bf c}}_2 \cap {\bar {\bf c}}_3$.

For $c \in {\bf c}_1 \cap {\bf c}_2 \cap {\bar {\bf c}}_3$,
we know from the consistency property that
\bear{fairK0012}
&&\pr \Big (\phi_1({\bf c}_1)=\phi_1({\bf c}_2)\ne  \phi_1({\bf c}_3) \Big |
\phi_1({\bf c}_0)=c\Big ) \nonumber\\
&&=\pr \Big (\phi_1({\bf c}_1)=\phi_1({\bf c}_2)=c,  \phi_1({\bf c}_3) \ne c \Big |
\phi_1({\bf c}_0)=c\Big ) \nonumber\\
&& =1.
\eear
Also, for $c \in {\bf c}_1 \cap {\bf c}_2 \cap {\bf c}_3$, we have from the consistency property that
the three users select $c$ and
\beq{fairK0013}
\pr \Big (\phi_1({\bf c}_1)=\phi_1({\bf c}_2)\ne  \phi_1({\bf c}_3)) \Big |
\phi_1({\bf c}_0)=c\Big )=0.
\eeq
Following the same argument, we also know that
the conditional probability is  0 for $c \in {\bar {\bf c}}_1 \cap {\bf c}_2 \cap {\bf c}_3$,
 $c \in {\bf c}_1 \cap {\bar {\bf c}}_2 \cap {\bf c}_3$, $c \in {\bar {\bf c}}_1 \cap {\bf c}_2 \cap {\bar {\bf c}}_3$, and $c \in {\bf c}_1 \cap {\bar {\bf c}}_2 \cap {\bar {\bf c}}_3$.

It remains to compute the conditional probability for
 $c \in {\bar {\bf c}}_1 \cap {\bar {\bf c}}_2 \cap {\bf c}_3$. For this, consider another fictitious user with the available channel set ${\bf c}_1 \cup {\bf c}_2$ and then use the law of total probability
 to derive  the following:
\bear{fairK0023}
&&\pr \Big (\phi_1({\bf c}_1)=\phi_1({\bf c}_2)\ne  \phi_1({\bf c}_3)) \Big |
\phi_1({\bf c}_0)=c\Big ) \nonumber\\
&&
=\sum_{c^\prime \in {\bf c}_1 \cup {\bf c}_2}
\pr \Big (\phi_1({\bf c}_1)=\phi_1({\bf c}_2)\ne  \phi_1({\bf c}_3))  \nonumber\\
&&\quad\quad\quad\quad \Big |
\phi_1({\bf c}_1 \cup {\bf c}_2)=c^\prime, \phi_1({\bf c}_0)=c\Big )\nonumber\\
&&\quad\quad\quad\quad \pr (\phi_1({\bf c}_1 \cup {\bf c}_2)=c^\prime |\phi_1({\bf c}_0)=c)
.
\eear
To compute \req{fairK0023}, we partition the set $c^\prime \in {\bf c}_1 \cup {\bf c}_2$ into two sets: $c^\prime \in {\bf c}_1 \cap {\bf c}_2$
and $c^\prime \not \in {\bf c}_1 \cap {\bf c}_2$.
Since $c \in {\bar {\bf c}}_1 \cap {\bar {\bf c}}_2 \cap {\bf c}_3$, we know from the consistency property that
user 3 selects c. Thus, for $c^\prime \in {\bf c}_1 \cap {\bf c}_2$, we know that $c^\prime \ne c$ and that
\bear{fairK0024}
&&\pr \Big (\phi_1({\bf c}_1)=\phi_1({\bf c}_2)\ne  \phi_1({\bf c}_3) \nonumber\\
&&\quad\quad\quad\quad\Big |
\phi_1({\bf c}_1 \cup {\bf c}_2)=c^\prime, \phi_1({\bf c}_0)=c\Big ) \nonumber\\
&&=\pr \Big (\phi_1({\bf c}_1)=\phi_1({\bf c}_2)=c^\prime ,\phi_1({\bf c}_3)=c) \nonumber\\
&&\quad\quad\quad\quad \Big |
\phi_1({\bf c}_1 \cup {\bf c}_2)=c^\prime, \phi_1({\bf c}_0)=c\Big ) \nonumber\\
&&=1 .
\eear
On the other hand, for $c^\prime \not \in {\bf c}_1 \cap {\bf c}_2$, we have from the consistent property that
\bear{fairk0024b}
&&\pr \Big (\phi_1({\bf c}_1)=\phi_1({\bf c}_2)\ne  \phi_1({\bf c}_3) \nonumber\\
&&\quad\quad\quad\quad\Big |
\phi_1({\bf c}_1 \cup {\bf c}_2)=c^\prime, \phi_1({\bf c}_0)=c\Big )\nonumber\\
&&=0.
\eear

Recall that $\phi_1=\pi_1^{-1} \circ \phi^{\rm min} \circ \pi_1$ with $\pi_1$ being a random permutation.
Using the symmetry property for a random permutation, one can follow the argument in the proof of  \rlem{fairgen} to show that for $c \in {\bar {\bf c}}_1 \cap {\bar {\bf c}}_2 \cap {\bf c}_3$ and $c_1 , c_2 \in {\bf c}_1 \cup {\bf c}_2$,
\bear{fairk0024c}
&&\pr (\phi_1({\bf c}_1 \cup {\bf c}_2)=c_1 , \phi_1({\bf c}_0)=c)\nonumber\\
&&=\pr (\phi_1({\bf c}_1 \cup {\bf c}_2)=c_2 , \phi_1({\bf c}_0)=c).
\eear
This then leads to
\beq{fairK0025}
\pr (\phi_1({\bf c}_1 \cup {\bf c}_2)=c^\prime |\phi_1({\bf c}_0)=c)= \frac{1}{|{\bf c}_1 \cup {\bf c}_2 |}.
\eeq
In view of \req{fairK0023}--\req{fairK0025}, we conclude that for $c \in {\bar {\bf c}}_1 \cap {\bar {\bf c}}_2 \cap {\bf c}_3$,
\bear{fairK0026}
\pr \Big (\phi_1({\bf c}_1)=\phi_1({\bf c}_2)\ne  \phi_1({\bf c}_3)) \Big |
\phi_1({\bf c}_0)=c\Big ) =\frac{|{\bf c}_1 \cap {\bf c}_2 |}{|{\bf c}_1 \cup {\bf c}_2 |}.
\eear

Using \req{fairK3344}, \req{fairK0012}, \req{fairK0013}, and \req{fairK0026}, we have from the seven cases for the partition of ${\bf c}_0$ in \req{fairK3333} that
\bear{fairK3366b}
&&\pr (E_{1,2,\bar 3})
=\frac{|{\bf c}_1 \cap {\bf c}_2 \cap {\bar {\bf c}}_3|}{|{\bf c}_1 \cup {\bf c}_2 \cup {\bf c}_3|}
+\frac{|{\bf c}_1 \cap {\bf c}_2 |}{|{\bf c}_1 \cup {\bf c}_2 |} \frac{|{\bar {\bf c}}_1 \cap {\bar {\bf c}}_2 \cap {\bf c}_3|}{|{\bf c}_1 \cup {\bf c}_2 \cup {\bf c}_3|}\nonumber\\
&&
=\frac{n_{1,2}-n_{1,2,3}}{n^{\rm union}}\nonumber\\
&&\quad+\frac{n_{1,2}}{n_1+n_2-n_{1,2}}\frac{n_3-n_{2,3}-n_{1,3}+n_{1,2,3}}{n^{\rm union}}.\nonumber\\
\eear

The derivations for \req{fairK5555} and \req{fairK6666} are similar.
Also,  \req{fairK7777} follows directly from \rlem{fairgen}.
Since the sum of these five events equals 1, we have \req{fairK7788}.
\eproof

Extending the analysis for $K \ge 4$ is significantly more challenging. This is because there is a nonzero probability that more than two groups of users may rendezvous at the same time (which is not possible for $K=3$).
Such a probability cannot be calculated by using the consistency property alone.

\bsubsec{The spread-out strategy for three users}{spread}

If the set of common channels is known to all users, each user can utilize this information to generate its channel hopping sequence, ensuring all users rendezvous within a single time slot. Motivated by this, we propose the \textbf{spread-out strategy} to gather the information of common channels.

We first consider the case with three users, i.e., $K=3$, in the synchronous setting. Assume that the sequence of permutations ${\bf \Pi}=\{\pi_1,\pi_2, \ldots\}$ in the ${\bf \Pi}$-algorithm is known to all users.
The process proceeds as follows:

\begin{enumerate}

    \item \textbf{Initialization}:
    Each user employs the $\mathbf{\Pi}$-algorithm along with its available channel set to generate its channel hopping sequence.

    \item \textbf{First Rendezvous} ($t_1$): Suppose at time $t_1$, two users, say user 1 and user 2 rendezvous.
    Upon rendezvous, they exchange information about their available channel sets. After this exchange, both users continue using their original channel hopping sequences.

    \item \textbf{Second Rendezvous} ($t_2$):
    Suppose one of these two users rendezvous with the third user at time $t_2$. Without loss of generality, assume users 1 and 3 rendezvous at $t_2$. After exchanging their available channel set information:
    \begin{itemize}
        \item Users 1 and 3 now know the available channel sets ${\bf c}_1$, ${\bf c}_2$, and ${\bf c}_3$ of all three users.
    \end{itemize}
    The remaining task is to ensure this information is passed to user 2.

    \item \textbf{Coordinated Information Passing}:
    Since ${\bf c}_2$ is known to users 1 and 3, they can predict the channel hopping sequence of user 2 after time $t_2$.
    \begin{itemize}
        \item User 1 computes the fastest rendezvous with user 2 using the $\mathbf{\Pi}$-algorithm with the available channel set ${\bf c}_1 \cap {\bf c}_2$.
        \item Similarly, user 3 computes the fastest rendezvous with user 2 using the $\mathbf{\Pi}$-algorithm with the available channel set ${\bf c}_3 \cap {\bf c}_2$.
    \end{itemize}
    Suppose user 1 is predicted to rendezvous with user 2 earlier than user 3. (In case of a tie, all three users will rendezvous simultaneously.) Then:
    \begin{itemize}
        \item User 1 is assigned the task of passing the full information to user 2.
        \item User 3 switches to using the $\mathbf{\Pi}$-algorithm with the reduced channel set ${\bf c}_1 \cap {\bf c}_2 \cap {\bf c}_3$ after $t_2$.
    \end{itemize}

    \item \textbf{Final Rendezvous} ($t_3$):
    Suppose user 1 meets user 2 at time $t_3$. After this rendezvous, user 2 now possesses the complete information about all available channel sets.
    \begin{itemize}
        \item Both user 1 and user 2 switch to using the $\mathbf{\Pi}$-algorithm with the reduced channel set ${\bf c}_1 \cap {\bf c}_2 \cap {\bf c}_3$.
        \item At time $t_3 + 1$, all three users will rendezvous.
    \end{itemize}

\end{enumerate}

\begin{figure}[!t]
    \centering
    \includegraphics[width=3in]{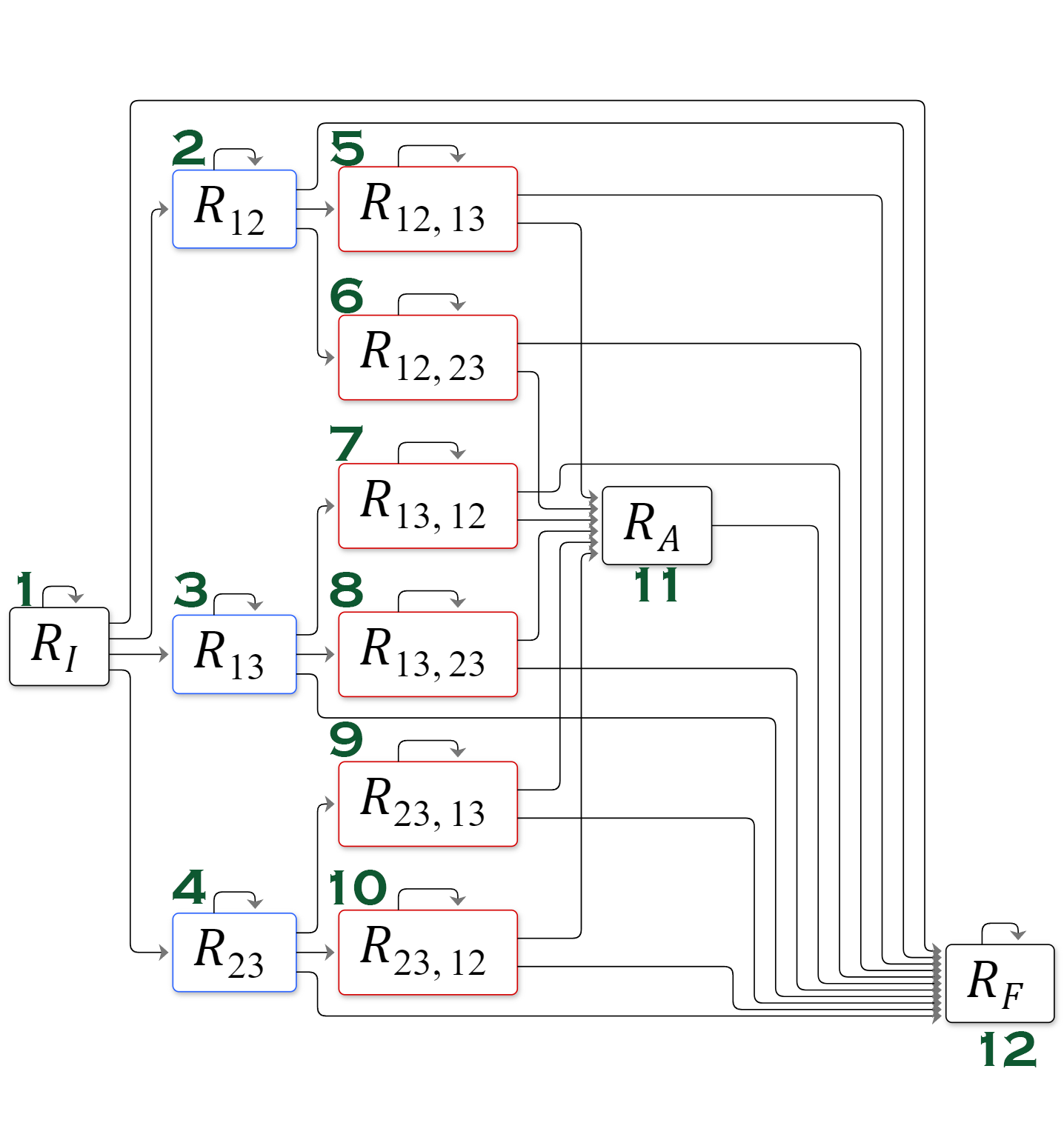}
    \caption{The state transition diagram for the rendezvous process with the spread-out strategy.}
    \label{fig:Spreadout State Diagram}
\end{figure}

Analogous to the Markov chain analysis for the stick-together strategy in \rsec{ETTR3}, we derive a closed-form expression for the ETTR of the spread-out strategy with three users, under the assumption that the ${\bf \Pi}$-algorithm shared by the three users is generated by (pseudo-)random permutations.
As in \rsec{ETTR3}, we model the rendezvous process by a discrete-time Markov chain \( \{X_0, X_1, X_2, \ldots \} \).
According to the spread-out strategy described in the previous section, there are 12 states of the Markov chain as shown in Fig. \ref{fig:Spreadout State Diagram}:
\begin{enumerate}
\item The initial state \( R_I \): where no rendezvous has yet occurred.
\item Three first rendezvous states \( R_{12}, R_{13}\) and \(R_{23} \): the state \( R_{12} \) represents that
 users 1 and 2 have rendezvoused. The same applies for \( R_{13} \) and \( R_{23} \). These three states are enclosed by blue rectangles in Fig. \ref{fig:Spreadout State Diagram}.
\item Six second rendezvous states \( R_{12,13}\), \(R_{12,23}\), \(R_{13,12}\), \(R_{13,23}\), \(R_{23,12}\), and \( R_{23,13} \): the state \( R_{12,13} \) represents that users 1 and 2 have already rendezvoused, followed by a rendezvous between users 1 and 3. The rest of the five states are defined similarly.  These six states are enclosed by red rectangles in Fig. \ref{fig:Spreadout State Diagram}. These represent all states in which one user has completed two pairwise rendezvous in sequence.
\item The mutual awareness state \( R_A \): the state represents all three users are aware of each other’s available channel set after coordinated information passing.
\item The final rendezvous state \( R_F \): the state represents all three users rendezvous together in the same time slot. This is also the absorbing state of the Markov chain.
\end{enumerate}

In this Markov chain, every state except \( R_A \) has a self-loop, indicating the possibility of remaining in the same state in the next time slot. Additionally, each state has a direct transition to the final state \( R_F \), representing the chance that all users may rendezvous spontaneously from any time.

The detailed derivation of the transition probabilities of the Markov chain is shown in Appendix A.

The ETTR is  the expected number of time slots required for the Markov chain to transition from the initial state \( R_I \) to the absorbing final rendezvous state \( R_F \), where all users have rendezvoused. To simplify notation, we relabel the original states \( R_I, R_{12}, \ldots, R_F \) using integer indices \( 1, 2, \ldots, 12 \). These integer states correspond to the green numbers annotated in Figure \ref{fig:Spreadout State Diagram}.

\begin{itemize}
    \item State 1: \( R_I \)
    \item States 2-4: \( R_{12}, R_{13}, R_{23} \)
    \item States 5-10: \( R_{12,13}, R_{12,23}, R_{13,12}, R_{13,23}, R_{23,12}, R_{23,13} \)
    \item State 11: \( R_A \)
    \item State 12: \( R_F \) (absorbing state).
\end{itemize}
Let \( p_{i,j} \) denote the transition probability \( \pr(X_{t+1} = j \mid X_t = i) \) from state \( i \) to state \( j \).
Note that the transition probabilities \( p_{i,j} \) are derived in 
\req{spread1111a}-\req{spread6666} 
in Appendix A.

Let \( T_i \) denote the absorption time from state \( i \), which represents the number of time slots required to reach the absorbing state \( 12 \) (corresponding to the final rendezvous state \( R_F \)) starting from state \( i \), i.e.,
\beq{spread11111}
T_i = \inf \left\{ t > 0 : X_t = 12 \,\middle|\, X_0 = i \right\}.
\eeq

Since the Markov chain transitions only forward, never returning to a less advanced state, we have
$p_{i,j} = 0 $ for all  $i > j$.
This enables recursive computation of the expected absorption time \(\ex [T_i] \). Specifically, for any transient state \( i < 12 \), the expected absorption time satisfies:
\beq{spreadettr}
\ex [T_i] = 1 + \sum_{j =i}^{11} p_{i,j} \cdot \ex [T_j].
\eeq
Here, the term \(1\) accounts for the current step, and the summation aggregates the expected times to all possible next states, weighted by their respective transition probabilities.

\bsubsec{A hybrid algorithm that alternates between the stick-together strategy and the spread-out strategy}{interleaving}

From the discussion in the previous section, we know that the spread‐out strategy is effective for collecting information about the common available channel set. Once every user knows the common available set, they can generate a channel hopping (CH) sequence that guarantees rendezvous in a single time slot. However, when there are more than three users, it is difficult to have all users reach consensus on the common available set. In view of this difficulty, we propose a hybrid algorithm that alternates between the stick‐together strategy and the spread‐out strategy on odd‐ and even‐numbered time slots, respectively.

\begin{enumerate}
    \item \textbf{Initialization}:
    Each user is given its available channel set and the same consistent channel hopping algorithm $\{\phi_1, \phi_2, \ldots\}$.

    User \(k\) maintains three sets:
    \begin{itemize}
        \item the available channel set \({\bf c}_k\),
        \item the known common channel set \({\bf c}_k^\prime\), and
        \item the stick‐together channel set \({\bf c}_k^{\prime\prime}\).
    \end{itemize}
    At \(t=1\), both \({\bf c}_k^\prime\) and \({\bf c}_k^{\prime\prime}\) are initialized to \({\bf c}_k\); that is,
    \[
    {\bf c}_k^\prime= {\bf c}_k^{\prime\prime}={\bf c}_k.
    \]

    \item \textbf{Sticking-together for odd-numbered time slots}:
    Suppose that \(t\) is an odd-numbered time slot. In this case, user \(k\) hops on the channel
    \[
    \phi_t({\bf c}_k^{\prime\prime}),
    \]
    meaning that the user selects a channel from its stick‐together channel set. If a group of users rendezvous at time \(t\), they exchange their known common channel sets. Then each user in the group updates its known common channel set by taking the intersection of all the exchanged sets. Finally, each user sets its stick‐together channel set equal to this updated known common channel set.

    \item \textbf{Spread-out for even-numbered time slots}:
    Suppose that \(t\) is an even-numbered time slot. In this case, user \(k\) hops on the channel
    \[
    \phi_t({\bf c}_k),
    \]
    meaning that the user selects a channel from its original available channel set. If a group of users rendezvous at time \(t\), they exchange their known common channel sets and update them by taking the intersection. However, in this case, the stick‐together channel set of each user remains unchanged.
\end{enumerate}

There are two key differences between the hybrid algorithm and the pure stick‐together strategy:
\begin{enumerate}
    \item For an even-numbered time slot, the channel that a user hops on is chosen from the available channel set \({\bf c}_k\) rather than from the stick‐together channel set \({\bf c}_k^{\prime\prime}\).
    \item For an even-numbered time slot, the stick‐together channel set \({\bf c}_k^{\prime\prime}\) is not updated; this ensures that every two time slots the users adhere to the stick‐together strategy.
\end{enumerate}

Denote by $T^{\rm hy}$ the TTR of the hybrid algorithm.
It is easy to see from the proof in \rthe{stick} that
$T^{\rm hy} \le T^{\rm uch}$ (the TTR for the strategy without changing the channel hopping sequences of the  $K$ users). However, whether $T^{\rm hy}$ is smaller $T^{\rm st}$ depends on the available channel sets of the $K$ users.

This hybrid design enables users to alternate between exploration (learning new intersections via spread-out moves) and exploitation (reinforcing consensus through stick-together moves), striking a balance that enhances convergence even when global consensus is hard to achieve.

\bsec{Using consistent channel hopping algorithms in the asynchronous setting}{asynset}

Now we extend the above algorithms to the asynchronous
 setting. Specifically, we consider the sym/async/hetero/global
 MRP. In such a setting, (i) sym: users are indistinguishable, (ii)
 async: users’ clocks may not be synchronized with the global
 clock, (iii) hetero: users might not have the same available
 channel sets, and (iv) global: the labels of the channels of a
 user are the same as the global labels of the channels.

Suppose that each user uses the same consistent channel hopping algorithm $\{\phi_1, \phi_2, \ldots, \}$. 
Similar to the dimension reduction technique in the LSH4 algorithm \cite{LSH}, each user generates a multiset of channel selection functions:
\[
\tilde{\bf \Phi} = \{\phi_1, \phi_2, \ldots, \phi_{T_0}\}.
\]
If the consistent channel hopping algorithm is generated by 
a sequence of pseudo-random permutations, we have from \rthe{fair} that the probability that none of the common channels between two users appear in their respective multisets is \((1 - J)^{T_0}\), where \(J\) is the Jaccard index of their available channel sets. This probability is negligible in practice when \(J > 0.5\) and \(T_0 = 20\), as shown in \cite{LSH}.

Essentially, the dimension reduction technique in the LSH4 algorithm \cite{LSH} can be generalized to the following procedure: at each time slot \(t\), each user either
\begin{itemize}
    \item with probability \(p_0\), selects a random index \(\tilde{t}\) uniformly in $[1, T_0]$, and uses the channel selection function \(\phi_{\tilde{t}}\) to select the channel;
    \item with probability \(1 - p_0\), uses the channel selection function \(\phi_t\) (corresponding to the current time) to select the channel.
\end{itemize}
The input to each channel selection function is dynamically determined by the stick-together or spread-out strategy described earlier.
Furthermore, after a successful rendezvous between two or more users, we require that they become synchronized. That is, once synchronized, the users should perform identical actions---selecting channels either based on the current time or from the shared multiset---thereby maintaining aligned behavior across future time slots.

To enable synchronization, we use seeded pseudo-random number generators to produce deterministic sequences that behave like sequences of uniformly distributed random variables. These generators are widely supported in most programming languages, including Python.

In the following, we outline the steps to
implement the hybrid strategy in the asynchronous setting.

\begin{enumerate}
    \item \textbf{Initialization: }
    Each user is given:
    \begin{itemize}
       \item A sequence of channel selection functions \(\{\phi_t\}_{t \ge 1}\), where each \(\phi_t(\cdot)\) is a consistent channel selection function,
        \item Two seeded pseudo-random number generators \( G_1, G_2 : \mathbb{Z} \times \mathbb{N} \to [0,1) \),
        \item Its available channel set \({\bf c}\).
      \end{itemize}
    At time \(t = 1\), each user randomly selects a \( \mathbf{seed} \in \mathbb{Z} \) for the two seeded pseudo-random number generators and sets
    \[
    {\bf c}^\prime = {\bf c}^{\prime\prime} = {\bf c},
    \]
where \({\bf c}^\prime\) is the known common channel set and \({\bf c}^{\prime\prime}\) is the stick-together channel set. The user keeps the state variables $\{t, \mathbf{seed}, {\bf c}, {\bf c}^\prime , {\bf c}^{\prime\prime}\}$.

    \item \textbf{Determining channel set \(c^\star\) at time $t$}:
    \[
    {\bf c}^\star =
    \begin{cases}
        {\bf c}^{\prime\prime}, & \text{if } t \text{ is odd}, \\
        {\bf c}, & \text{if } t \text{ is even}.
    \end{cases}
    \]

    \item \textbf{Channel selection at time $t$:}
    \[
    p = G_1(\mathbf{seed}, t), \quad
    t' = \left\lfloor T_0 \cdot G_2(\mathbf{seed}, t) \right\rfloor + 1,
    \]

    \begin{itemize}
        \item If \(p < p_0\), hop to \(\phi_{t'}({\bf c}^{\star})\).
        \item Otherwise, randomly select a channel from \({\bf c}^\star\) and hop to the channel.

    \end{itemize}
    \item \textbf{Update:}
       Upon rendezvous of a group of users, each user in the group:
\begin{itemize}
    \item Arbitrarily selects a common reference user from the group;
    \item Synchronizes its clock \(t\) and seed \(\mathbf{seed}\) to that of the reference user;
    \item Updates \({\bf c}^\prime\) as the intersection of the known common channel sets of all users in the group;
    \item If \(t\) is odd, also sets \({\bf c}^{\prime\prime} \leftarrow {\bf c}^\prime\).
\end{itemize}

\end{enumerate}

The stick-together strategy can be viewed as a special case of the hybrid strategy, where users always hop using their stick-together sets \({\bf c}^{\prime\prime}\) at every time slot. Upon each rendezvous, users exchange their known common sets and update both their known common set \({\bf c}^{\prime}\) and stick-together set \({\bf c}^{\prime\prime}\) to the intersection of all exchanged sets.

\bsec{Simulations}{sim}

\bsubsec{Two users}{two}

Our experiments are designed similarly to those described in \cite{LSH}, and we use this design to generate the available channel sets for the two users.  For each experiment, we simulate 20,000 time slots and record the TTR. We conduct 10,000 experiments and estimate the ETTR by averaging the TTRs from these 10,000 experiments.
To calculate the MTTR, we find the maximum TTR in every batch of 100 experiments and then average these maximum values over the 100 batches.
In all our experiments, we set $N=256$ and $n_1=n_2=60$.

\begin{figure}[!t]
	\centering
	\includegraphics[width=2.5in]{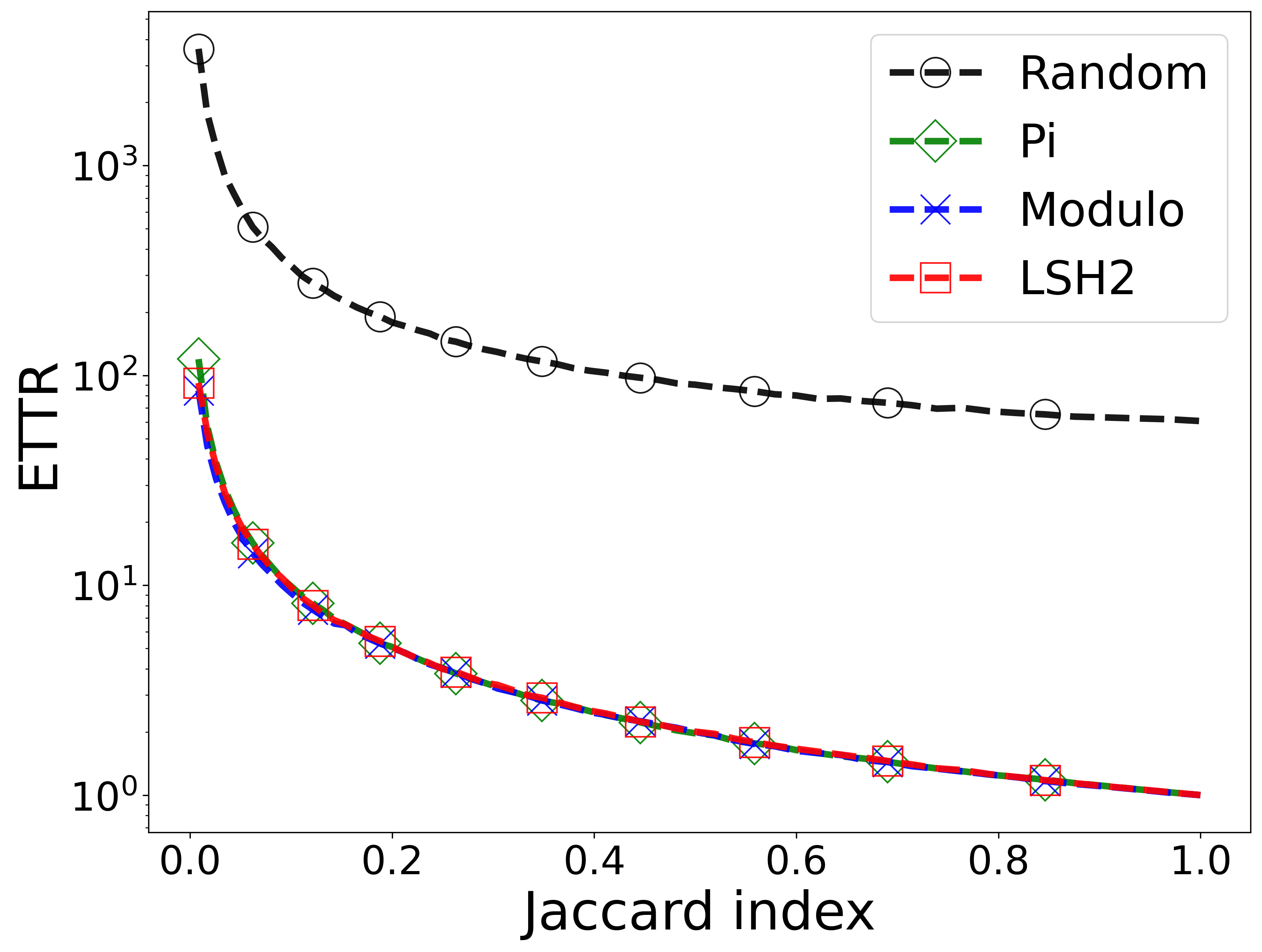}
	\caption{The ETTR in the synchronous setting with two users.}
	\label{fig:comparelslsh2ETTR}
\end{figure}

In \rfig{comparelslsh2ETTR}, we first compare the ETTR of the $\bf \Pi$-algorithm with random permutations (marked with Pi), the modulo algorithm (marked with Modulo), the LSH2 algorithm \cite{LSH} (marked with LSH2) and  the random algorithm (marked with Random)  when the clocks of the two users are synchronized. For the modulo algorithm, we select the prime \(P = 257\) and choose the generator \(g = 254\). Note that  the ETTR of the $\bf \Pi$-algorithm with random permutations, the modulo algorithm and  the LSH2 algorithm \cite{LSH} approach $1/J$ as shown in \rthe{fair}.

\begin{figure}[!t]
	\centering
	\includegraphics[width=2.5in]{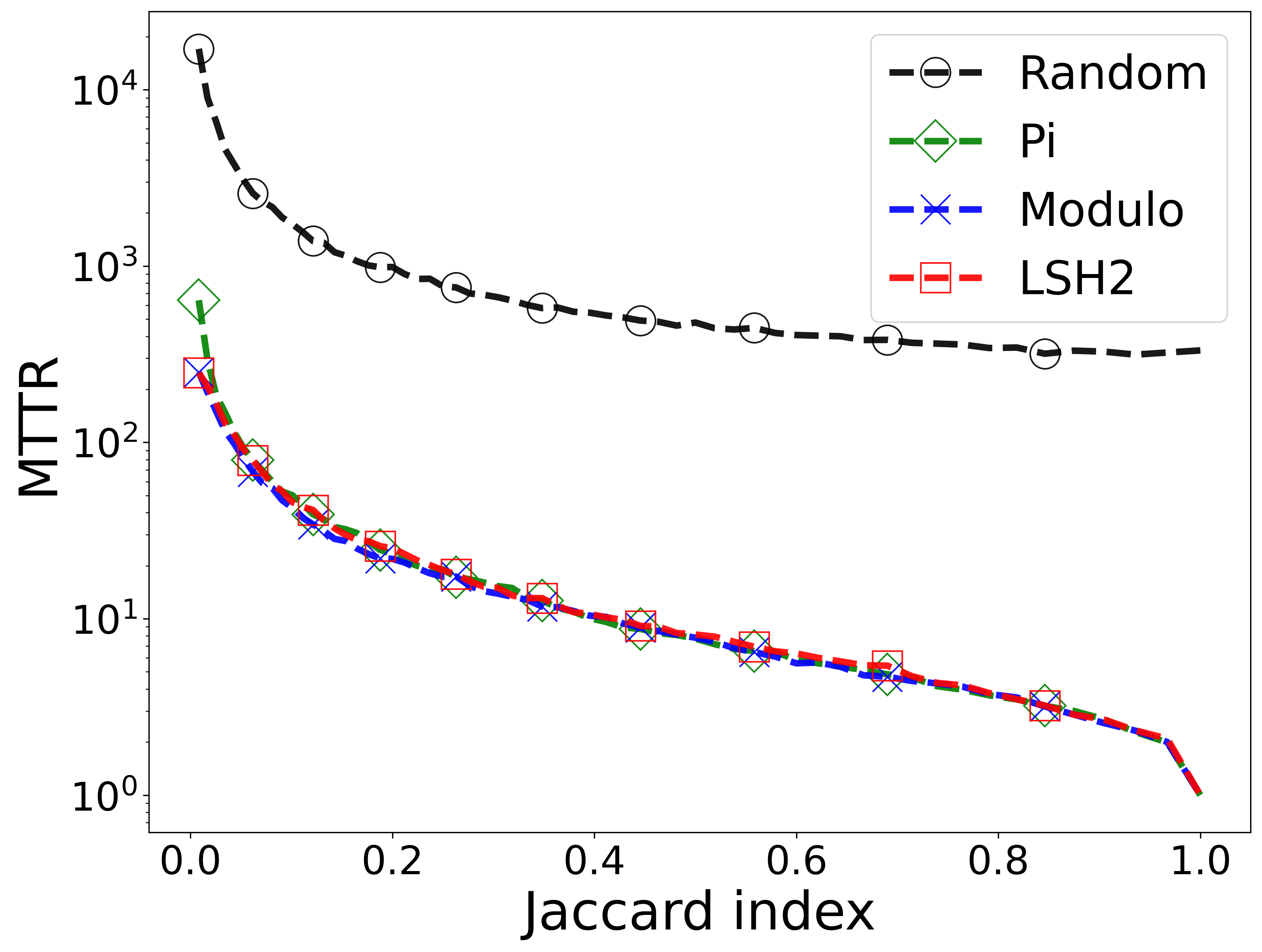}
	\caption{The MTTR in the synchronous setting with two users.}
	\label{fig:comparelslsh2MTTR}
\end{figure}

Moreover, as shown in \rfig{comparelslsh2MTTR}, the MTTR of the modulo algorithm is also close to that of the LSH2 algorithm. Notably, the ETTR and MTTR performance metrics of both the LSH2 and the modulo algorithms surpass those of the random algorithm.

\begin{figure}[!t]
	\centering
	\includegraphics[width=2.5in]{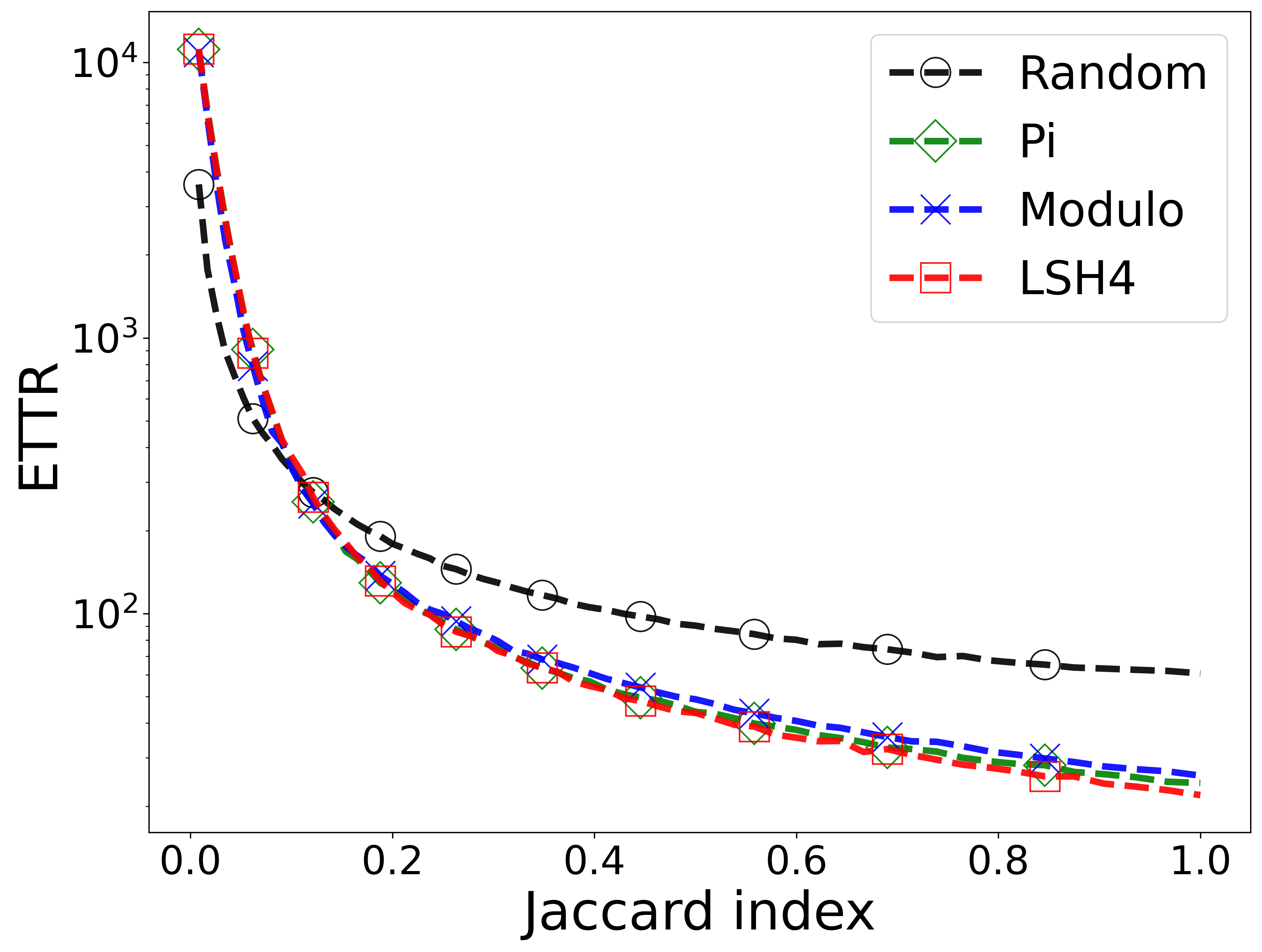}
	\caption{The ETTR in the asynchronous setting with two users.}
	\label{fig:comparelslsh4ETTR}
\end{figure}

\begin{figure}[!t]
	\centering
	\includegraphics[width=2.5in]{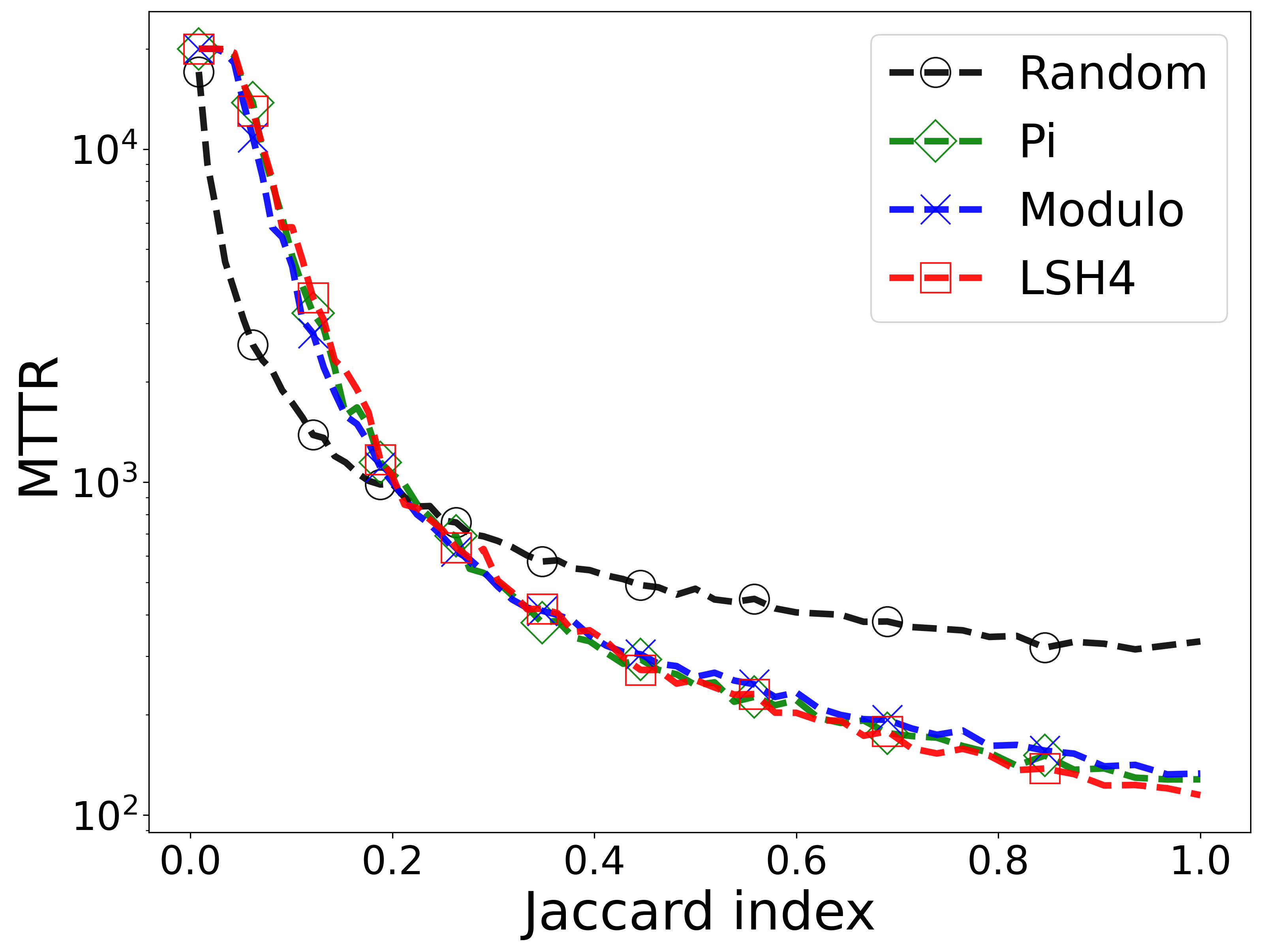}
	\caption{The MTTR in the asynchronous setting with two users.}
	\label{fig:comparelslsh4MTTR}
\end{figure}

We then compare the modulo algorithm  with the random algorithm and the LSH4 algorithm \cite{LSH} when the clocks of the two users are not synchronized. For both the modulo algorithm and the LSH4 algorithm, we use the parameter settings $T_0=20$ and $p_0=0.75$.
In \rfig{comparelslsh4ETTR} and \rfig{comparelslsh4MTTR}, we find that the ETTR and the MTTR of the modulo algorithm
in the asynchronous setting  closely match those of the LSH4 algorithm and they are better than those of the random algorithm for $J \ge 0.3$.

\bsubsec{Three users}{three}

The experiments with three users follow a design analogous to the experiments with two users. In these simulations, we set the total number of channels to \(N = 256\) and assume that each user has 60 available channels i.e., \(n_1 = n_2 = n_3 = 60\).  

To generate the available channel sets for the three users, we randomly partition
the channel set \(\{1, 2, \ldots, N\}\) into eight disjoint subsets with the following sizes:
\begin{description}
\item[-] \(n_{1,2,3}\) (shared by all three users),
\item[-] \(n_{1,2} - n_{1,2,3}\) (shared by users 1 and 2 only),
\item[-] \(n_{1,3} - n_{1,2,3}\) (shared by users 1 and 3 only),
\item[-] \(n_{2,3} - n_{1,2,3}\) (shared by users 2 and 3 only),
\item[-] \(n_1 - n_{1,2} - n_{1,3} + n_{1,2,3}\) (exclusive to user 1),
\item[-] \(n_2 - n_{1,2} - n_{2,3} + n_{1,2,3}\) (exclusive to user 2),
\item[-]  \(n_3 - n_{1,3} - n_{2,3} + n_{1,2,3}\) (exclusive to user 3),
\item[-] and the remaining unused channels.
\end{description}
The available channels of each user are then constructed from
the union of the subsets corresponding to its specific sharing pattern.

Let \( n_{\mathrm{core}} \) denote the number of channels commonly available to all three users, i.e.,
\beq{stm1112}
n_{\mathrm{core}} = n_{1,2,3} = \left| {\bf c}_1 \cap {\bf c}_2 \cap {\bf c}_3 \right|.
\eeq
We refer to the channels in ${\bf c}_1 \cap {\bf c}_2 \cap {\bf c}_3$ as the \emph{core channels}.

Let \( n_{\mathrm{exclusive}}^{(1,2)} \) denote the number of channels shared by users 1 and 2 but not user 3, i.e.,
\beq{stm1111}
n_{\mathrm{exclusive}}^{(1,2)} = n_{1,2} - n_{\mathrm{core}} = \left| {\bf c}_1 \cap {\bf c}_2 \right| - \left| {\bf c}_1 \cap {\bf c}_2 \cap {\bf c}_3 \right|.
\eeq
We refer to these channels as the \emph{pairwise-exclusive channels} between user 1 and user 2.

In addition, we assume that the sizes of all pairwise-exclusive channels are the same, i.e.,
\beq{stm1111b}
n_{\mathrm{exclusive}}=n_{\mathrm{exclusive}}^{(1,2)} = n_{\mathrm{exclusive}}^{(1,3)} = n_{\mathrm{exclusive}}^{(2,3)}.
\eeq

From \req{stm1111b}, we know that the number of channels exclusive to user 1 is
$n_1-2n_{\mathrm{exclusive}}- n_{\mathrm{core}}$.
Since $n_1=n_2=n_3=60$, the numbers of channels exclusive for the three users are the same.
Therefore, our experiment involves only two control variables: \( n_{\mathrm{exclusive}} \) and \( n_{\mathrm{core}} \). As noted in \req{stm1111}, when \( n_{\mathrm{exclusive}} \) is small, \( n_{\mathrm{core}} \) is close to \( n_{1,2} \), making it likely that all three users will rendezvous simultaneously. In contrast, when \( n_{\mathrm{exclusive}} \) is large, rendezvous events are more likely to occur between user pairs rather than among all three users at once.

\bsubsubsec{Three users in the synchronous setting}{3sync}

    \begin{figure}[!ht]
        \centering
        \includegraphics[width=2.5in]{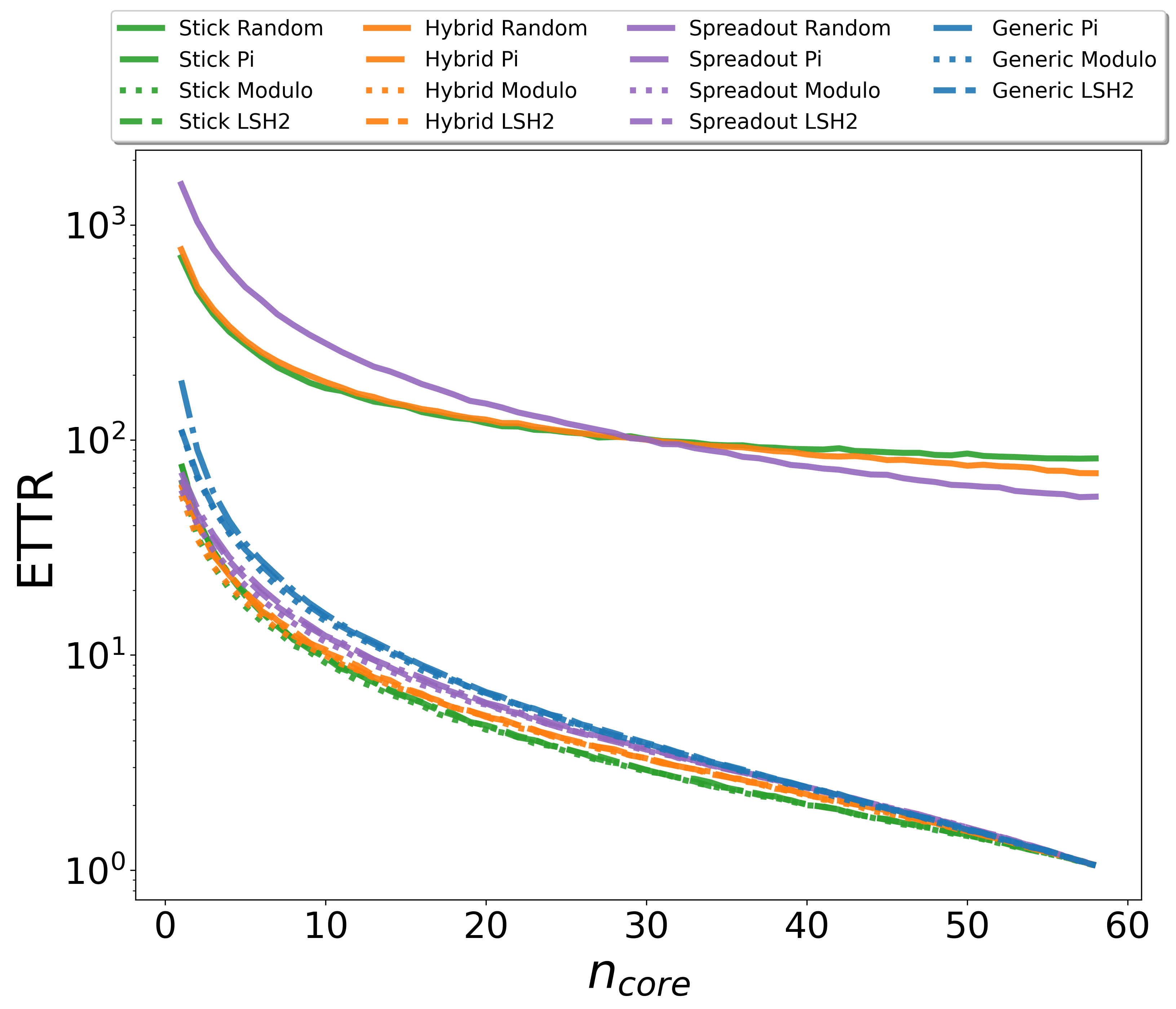}
        \caption{The ETTR  (as a function of $n_{\mathrm{core}}$) in the synchronous setting with three users and $n_{\mathrm{exclusive}}=1$.}
        \label{fig:3user-ettr-fixed-x2}
    \end{figure}

In this experiment, we consider the rendezvous problem in the synchronous setting with three users and a fixed number of pairwise-exclusive channels \( n_{\mathrm{exclusive}} = 1 \). Since \( n_{\mathrm{exclusive}}\) is only one,
$n_{\mathrm{core}}$ is very close to $n_{1,2}$.
In \rfig{3user-ettr-fixed-x2}, we compare the ETTR (as a function of $n_{\mathrm{core}}$) of
the $\bf \Pi$-algorithm with random permutations (marked with Pi), the modulo algorithm (marked with Modulo), the LSH2 algorithm \cite{LSH} (marked with LSH2) and  the random algorithm (marked with Random) under various strategies:
the direct extension in \rsec{direct} (marked with Generic), the stick-together strategy in \rsec{stick} (marked with Stick), the spread-out strategy in \rsec{spread} (marked with Spreadout), and the hybrid strategy in \rsec{interleaving} (marked with Hybrid). Note that the random algorithm in this experiment uses the seeded pseudo-random number generators (as mentioned in \rsec{asynset}) so that two rendezvous users can exchange their seeds to predict future channel selections of these two users.   There are several findings of this experiments.
\begin{description}
\item[(i)] As in the case with two users, the three consistent channel hopping algorithms, i.e., the $\bf \Pi$-algorithm with random permutations,
the modulo algorithm, and  the LSH2 algorithm, have comparable ETTR performance under each of the four strategies.
\item[(ii)]
Under the same strategy, the performance of
the random algorithm is much worse than any one of the three consistent channel hopping algorithms.
\item[(iii)] Since \( n_{\mathrm{exclusive}} \) is very small, the stick-together strategy achieves the best performance due to its rapid convergence once two users have rendezvoused. However, the gain is rather small, compared to the hybrid and spread-out strategies. Moreover, as illustrated in our next experiment (see \rfig{3user-ettr-fixed-x3} below),  the stick-together strategy becomes less effective as  \( n_{\mathrm{exclusive}} \) increases.
\end{description}

    \begin{figure}[!ht]
        \centering
        \includegraphics[width=2.5in]{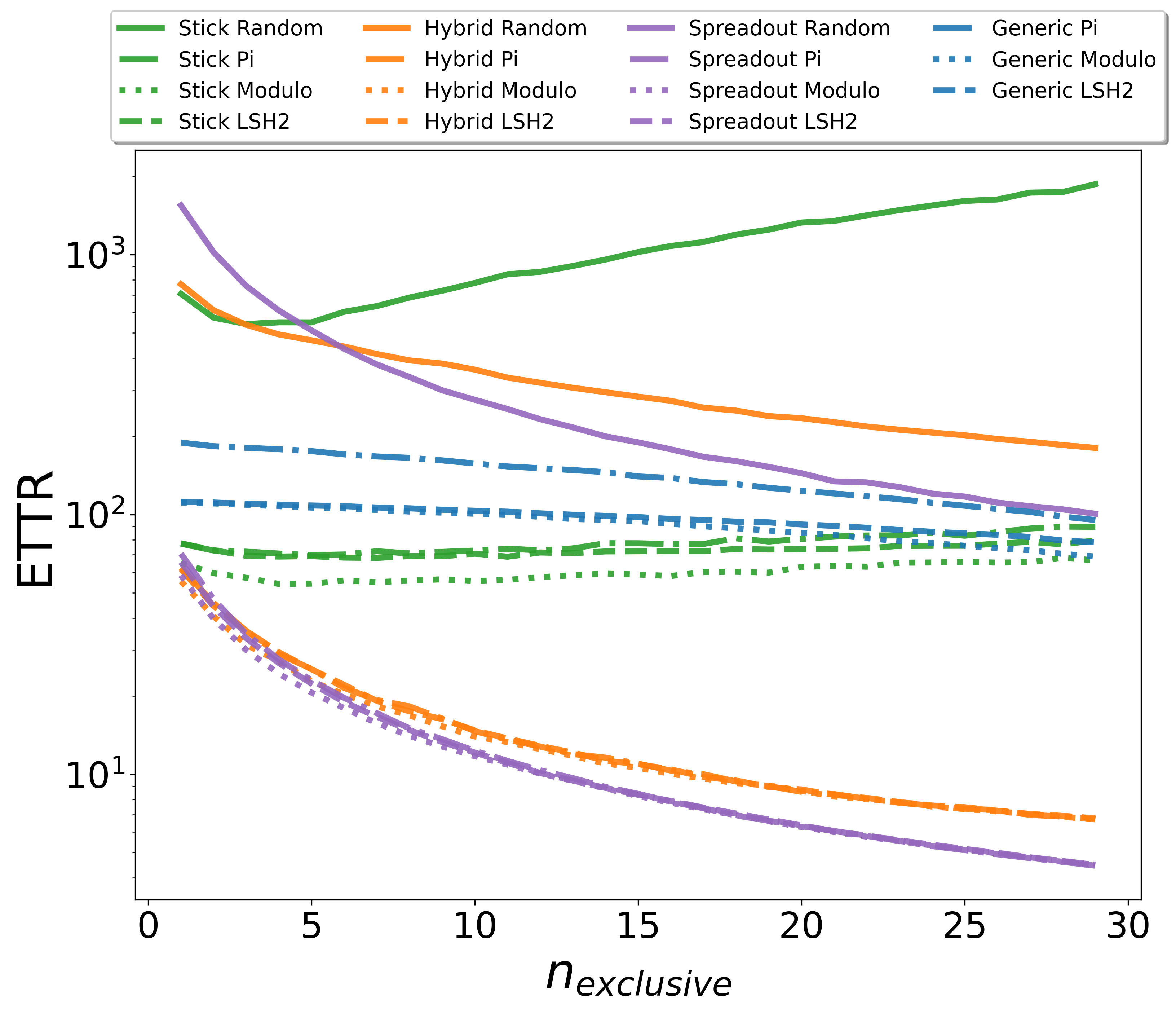}
        \caption{The ETTR (as a function of $n_{\mathrm{exclusive}}$) in the synchronous setting with three users and $n_{\mathrm{core}}=1$.}
        \label{fig:3user-ettr-fixed-x3}
    \end{figure}

In \rfig{3user-ettr-fixed-x3}, with a fixed value of \( n_{\mathrm{core}} = 1 \),
we compare the ETTR (as a function of $n_{\mathrm{exclusive}}$) of these algorithms and strategies.
Both the spread-out strategy and the hybrid strategy demonstrate the best performance. This is because the generic and stick-together strategies rely solely on the only common channel to rendezvous (\( n_{\mathrm{core}} = 1 \)). Also, the stick-together strategy suffers from a significant limitation: when only two users have rendezvoused, their available channel set becomes severely restricted (to the only common channel), which significantly reduces the chance of successfully rendezvousing with the third user.
By contrast, the spread-out and hybrid strategies are capable of effectively utilizing the pairwise rendezvous to maintain better connectivity and increase rendezvous opportunities when  \( n_{\mathrm{exclusive}} \) is large.

	\bsubsubsec{Three users in the asynchronous setting}{3async}

	\begin{figure}[!ht]
		\centering
		\includegraphics[width=2.5in]{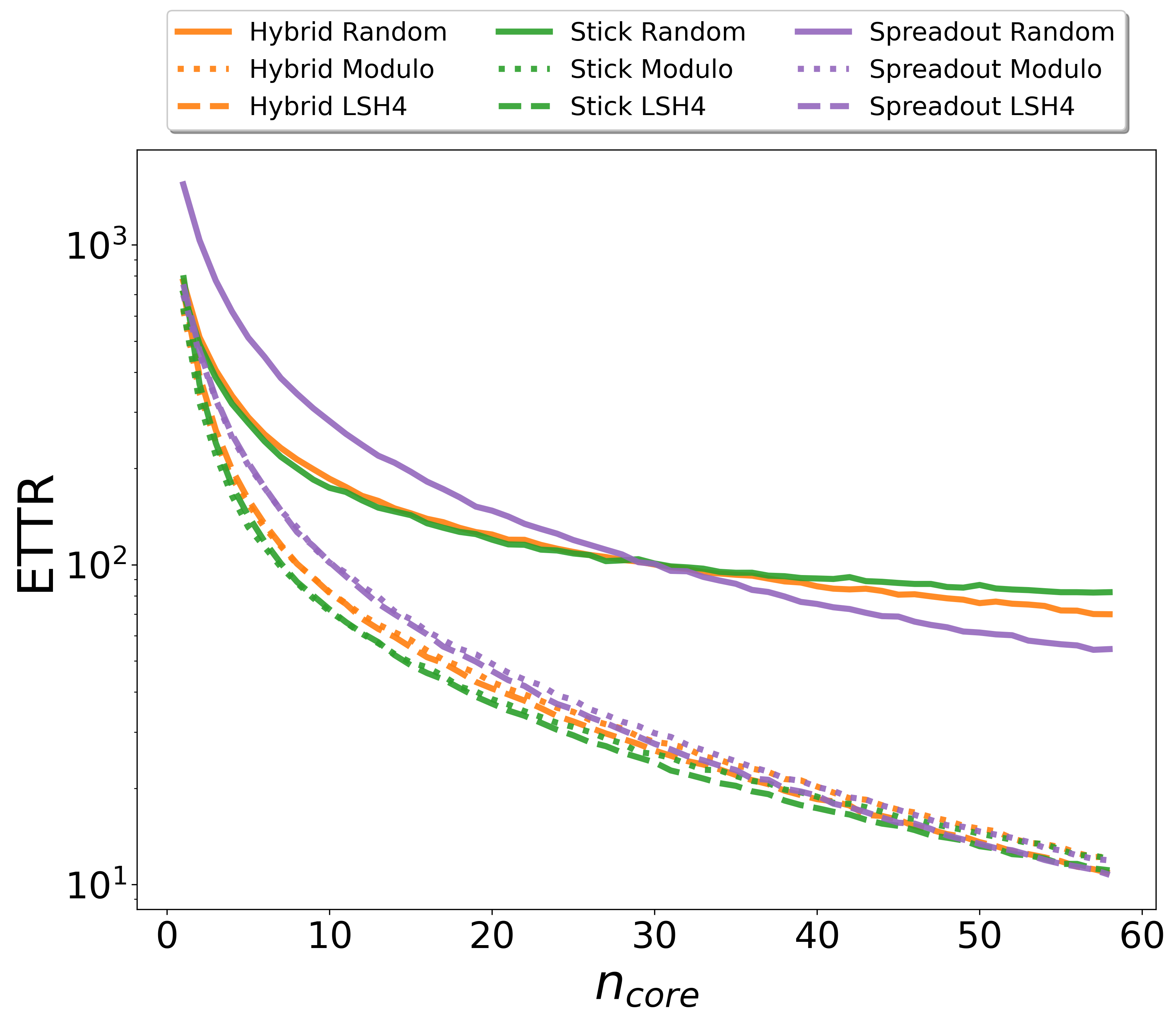}
        \caption{The ETTR of 3-user rendezvous in the asynchronous setting with $n_{\mathrm{exclusive}}=1$.}
		\label{fig:3usersETTRasynchronous2}
	
	\end{figure}
    \begin{figure}[!ht]
		\centering
        \includegraphics[width=2.5in]{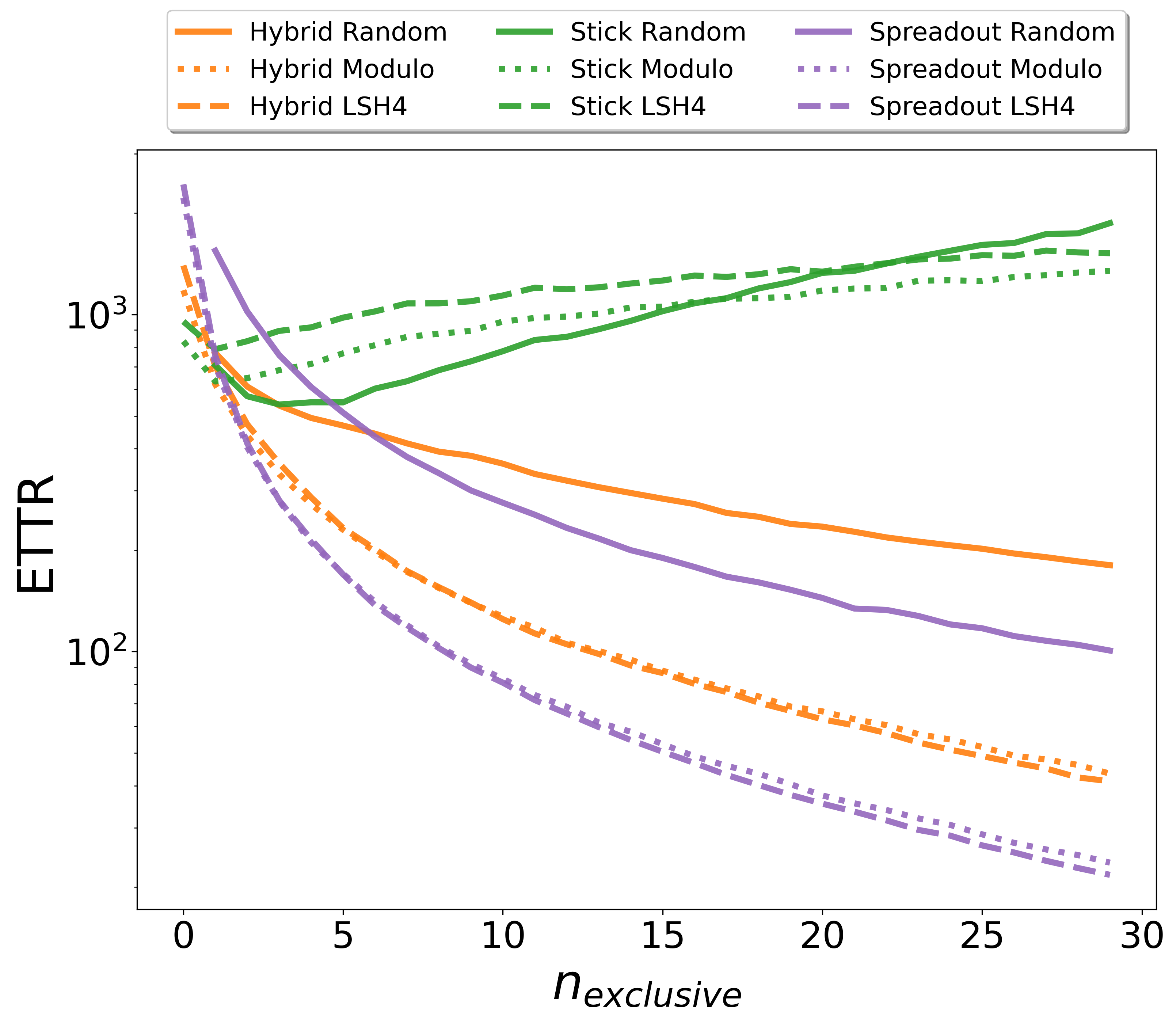}

        \caption{The ETTR of 3-user rendezvous in the asynchronous setting with $n_{\mathrm{core}}=1$.}
        \label{fig:3usersETTRasynchronous1}
    \end{figure}

We then evaluate the performance of these algorithms in the asynchronous setting. In these experiments, the parameters are set to \( T_0 = 20 \) and \( p_0 = 0.75 \). \\

In \rfig{3usersETTRasynchronous2} and \rfig{3usersETTRasynchronous1}, we compare the ETTR for various algorithms and strategies as described in the synchronous setting in \rsec{3sync}. As shown in these two figures, the performance of the modulo algorithm (an implementation of the ${\bf \Pi}$-algorithm with one-cycle permutations) is comparable to that of the LSH4
algorithm \cite{LSH}, which is consistent with the findings in the synchronous setting.
The performance of the modulo algorithm in the asynchronous setting is worse than in the synchronous setting (see \rfig{3user-ettr-fixed-x2}), as additional time is required for users to rendezvous and synchronize their clocks.
On the other hand, the performance of the random algorithm remains the same in both the asynchronous and synchronous settings. Despite this, the performance of the modulo algorithm  still
 outperforms the random algorithm under the same strategy, provided that the number of common channels---either among all users or between pairs of users---is not too small.
 In particular,  this is observed when \(n_{\mathrm{core}} > 5\) in \rfig{3usersETTRasynchronous2} and  when \(n_{\mathrm{exclusive}} > 5\) in \rfig{3usersETTRasynchronous1}. These results further confirm that consistent channel hopping algorithms remain effective even in the asynchronous setting.

In \rfig{3usersETTRasynchronous1}, we observe that both the spread-out strategy and the hybrid strategy demonstrate the best performance.
This aligns with the findings in the synchronous setting.
We also note that
the ETTR curve under the stick-together strategy initially decreases and then increases as \(n_{\mathrm{exclusive}}\) grows. This nonmonotonic trend arises from the way  the known common channel set \({\bf c}^{\prime}\) evolves during the process. When \(n_{\mathrm{exclusive}}\) is moderately large, early rendezvous events enable users to shrink \({\bf c}^{\prime}\) quickly. However, as \(n_{\mathrm{exclusive}}\) becomes too large, \({\bf c}^{\prime}\) remains large for longer periods. Selecting from a large \({\bf c}^{\prime}\), either directly or via a multiset, results in a lower rendezvous probability. As a result, the advantage of early synchronization diminishes, and the consistency property of the \(\Pi\)-algorithms, which typically increases the chance of rendezvous, becomes a liability, leading to an increase in ETTR for the stick-together strategy.

\bsubsec{A large number of users}{hundread}

In this section, we evaluate the performance of the consistent channel hopping algorithm for a large number of users. Specifically, we consider a cognitive radio network with primary users (PUs) and secondary users (SUs), where SUs can only use channels that are not interfered with by PUs.

We set the number of channels to \( N = 256 \) and the number of SUs to \( K = 100 \). To generate the network, we randomly distribute these 100 SUs within a square of dimensions \( 1{,}000 \) m \( \times \) \( 1{,}000 \) m.
We randomly select a subset of common channels, denoted by \( {\bf c}_{\rm core} \). The set of channels allocated to PUs is then defined as
\[
{\bf c}_{\mathrm{PU}} = \{1, \ldots, N\} \backslash {\bf c}_{\mathrm{core}}.
\]
The number of PUs is set to 50, and these PUs are randomly distributed within the same square area as the SUs. The interference range of a PU to an SU is set to \( 500 \) meters. If a PU has no SUs within its interference range, it does not interfere with SU communication and is therefore removed from the set of active PUs. Let \( P \) be the number of remaining PUs after this filtering process. The channels in \( {\bf c}_{\mathrm{PU}} \) are then assigned to these \( P \) PUs as evenly as possible. A PU is considered an \emph{interfering PU} to an SU if the SU falls within the PU's interference range. The available channel set for an SU consists of the remaining channels from the total \( N \) channels that are not occupied by its interfering PUs. Under this channel assignment strategy, the set of common channels available to all \( K \) SUs is exactly \( {\bf c}_{\mathrm{core}} \).

\begin{figure}[!ht]
		\centering
        \includegraphics[width=2.5in]{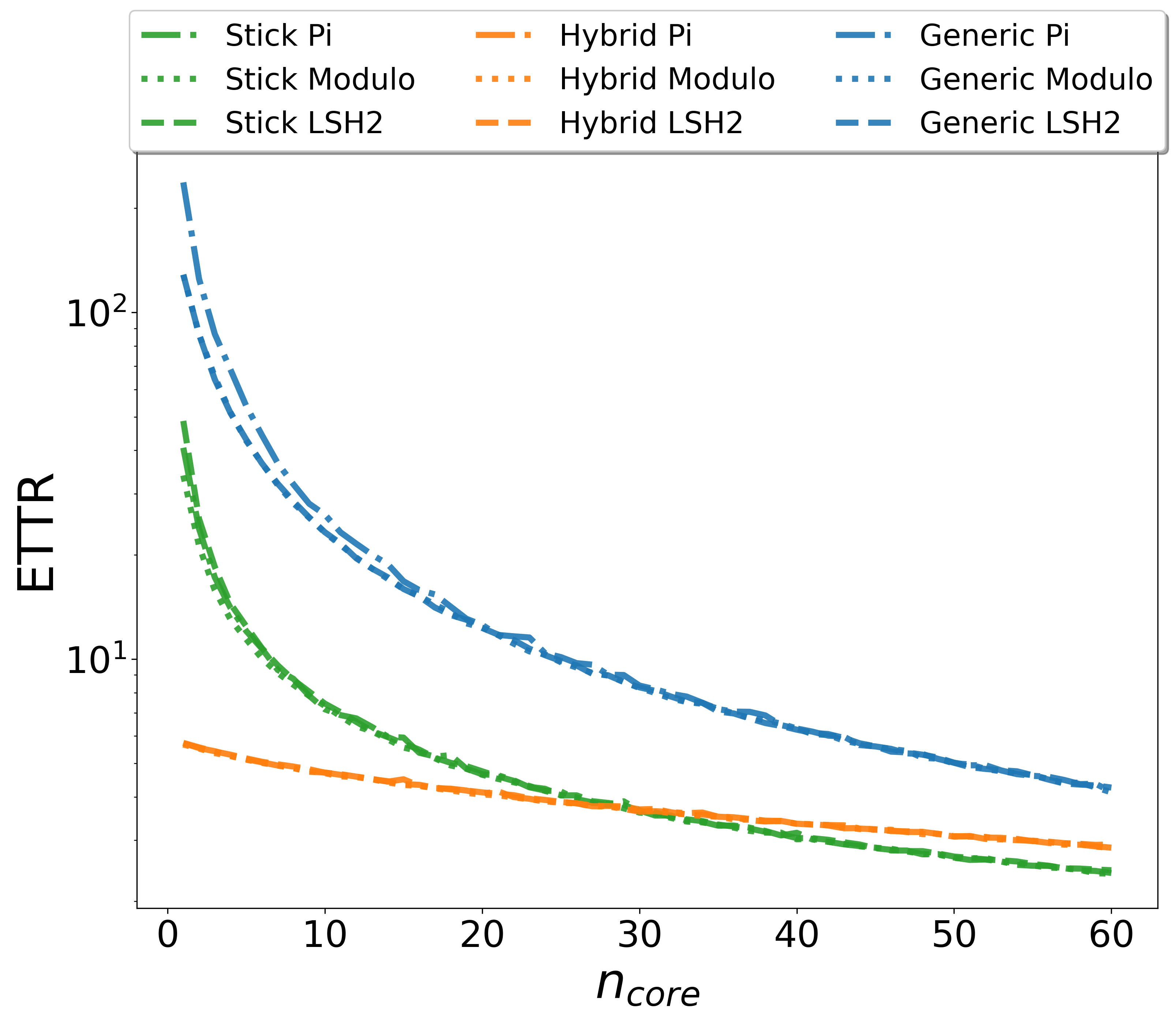}

        \caption{The ETTR of 100-user rendezvous in the synchronous setting.}
        \label{fig:100usersETTRsynchronous1}
    \end{figure}

In \rfig{100usersETTRsynchronous1}, we compare the ETTR of this rendezvous problem in the synchronous setting under three different strategies: generic, stick-together, and hybrid. As shown in the figure, the hybrid strategy performs consistently well across the entire range of \( {\bf c}_{\mathrm{core}} \). This is because the probability of pairwise rendezvous in each time slot increases with the number of users. As more users participate, the likelihood that any given user encounters another who already possesses partial or complete knowledge of the global intersection \( {\bf c}_{\mathrm{core}} \) increases rapidly. The hybrid strategy takes advantage of this by alternating between spread-out phases, which maximize coverage and discovery, and stick-together phases, which facilitate information propagation and convergence. Consequently, knowledge of the common channel set spreads exponentially faster in larger populations, reducing the time required for all users to become mutually aware. This collective acceleration effect makes the hybrid strategy particularly effective at scale.

\bsec{Conclusion}{con}

In this paper, we developed a theoretical framework for consistent channel hopping algorithms to solve the MRP in wireless networks with heterogeneous available channel sets. By formalizing the notion of consistency in channel selection functions, we showed that all consistent algorithms are functionally equivalent to selecting the smallest-index channel after appropriate relabeling. This led to a unifying representation of consistent channel hopping algorithms as permutation-based sequences.

We derived tight upper bounds on the MTTR using one-cycle permutations and proved that the ETTR for randomly generated permutations equals the inverse of the Jaccard index between available channel sets. Furthermore, we established the optimality of consistent algorithms in maximizing rendezvous probability at each time slot.

To reduce computational overhead, we proposed the modulo algorithm, which uses modular arithmetic to efficiently generate one-cycle permutations. The modulo algorithm achieves ETTR and MTTR performance comparable to that of more complex LSH-based algorithms while maintaining much lower computational complexity.

We extended our framework to the MRP with multiple users and proposed the stick-together, spread-out, and hybrid strategies to accelerate the rendezvous process. For the asynchronous setting, we generalized the dimension reduction technique in LSH-based methods to enable synchronization and efficient rendezvous using pseudo-random generators.

Extensive simulations confirm that consistent channel hopping algorithms, particularly the modulo and hybrid strategies, offer robust and scalable solutions to the MRP across a range of user configurations and system settings. These results establish a strong theoretical and practical foundation for future work on efficient rendezvous protocols in dynamic multichannel wireless environments.


\newpage
	\appendices

	\section{Derivation of  the transition probabilities of the Markov chain of spread-out strategy with three users}

In this appendix we derive the transition probabilities of the Markov chain of spread-out strategy with three users.

\begin{enumerate} 
    \item Transitions from the initial state \(R_I\):

At each time slot \( t \), exactly one of the following five events can occur from the initial state \(R_I\):
\begin{itemize}
    \item \( E_{1,2,\bar{3}} \): Users 1 and 2 rendezvous, but user 3 does not.
    \item \( E_{1,\bar{2},3} \): Users 1 and 3 rendezvous, but user 2 does not.
    \item \( E_{\bar{1},2,3} \): Users 2 and 3 rendezvous, but user 1 does not.
    \item \( E_{1,2,3} \): All three users rendezvous simultaneously.
    \item \( E_0 \): No users rendezvous.
\end{itemize}

Clearly, if $E_0$ occurs, the Markov chain remains in \(R_I\). If \( E_{1,2,3} \) occurs, then it transitions to the final state $R_F$. If \( E_{1,2,\bar{3}} \) (resp. \( E_{1,\bar{2},3} \), \(E_{\bar{1},2,3} \)) occurs, then it transitions into the first rendezvous state $R_{12}$ (resp. $R_{13}$, $R_{23}$). Thus, the transition probabilities from the initial state $R_I$ can be computed as follows:

\begin{align}
\pr(X_{t+1} = R_I \mid X_t = R_I) &= \pr(E_0 ), \label{eq:spread1111a}\\
\pr(X_{t+1} = R_{12} \mid X_t = R_I) &= \pr(E_{1,2,\bar{3}}), \\
\pr(X_{t+1} = R_{13} \mid X_t = R_I) &= \pr(E_{1,\bar{2},3} ), \\
\pr(X_{t+1} = R_{23} \mid X_t = R_I) &= \pr(E_{\bar{1},2,3} ), \\
\pr(X_{t+1} = R_F \mid X_t = R_I) &= \pr(E_{1,2,3} ),
\label{eq:spread1111}
\end{align}
where
the probabilities of these five events are derived in \rlem{five}.

\item Transition from the first rendezvous states:

Once two users have rendezvoused (i.e., the Markov chain is in state \( R_{12} \), \( R_{13} \), or \( R_{23} \)), all users continue to follow their original channel hopping sequences based on the same consistent selection function \( \phi_t \). Therefore, the event probabilities remain unchanged and are equal to those in the initial state \( R_I \).

Specifically, the transition probabilities from \( R_{12} \) are given by:
\begin{align}
    \pr(X_{t+1} = R_{12,13} \mid X_t = R_{12}) &= \pr(E_{1,\bar{2},3} ), \\
    \pr(X_{t+1} = R_{12,23} \mid X_t = R_{12}) &= \pr(E_{\bar{1},2,3} ), \\
    \pr(X_{t+1} = R_F \mid X_t = R_{12})        &= \pr(E_{1,2,3} ), \\
    \pr(X_{t+1} = R_{12} \mid X_t = R_{12})     &= \pr(E_{1,2,\bar{3}} ) + \pr(E_0 ).
    \label{spread2222}
\end{align}
The transition probabilities from \( R_{13} \) and \( R_{23} \) can be derived similarly.

\item Transition probabilities from the second rendezvous states:

Consider the state \( R_{12,13} \), where user 1 has met user 2 and subsequently met user 3. At this point, users 1 and 3 have exchanged information and are both aware of user 2’s available channel set \( {\bf c}_2 \). Since the channel hopping algorithm of each user follows the same ${\bf \Pi}$-algorithm with the same sequence of pseudo-random permutations, users 1 and 3 can predict user 2's future channel selections. As such, the transition at time \( t+1 \) depends solely on the channel selected by user 2.

Let \( \phi_{t+1}({\bf c}_2) \) denote the channel that user 2 hops to at time \( t+1 \).
There are three possible outcomes for the channel selected by user 2.

\begin{itemize}
    \item \( \phi_{t+1}({\bf c}_2) \in {\bf c}_1 \cap {\bf c}_2 \cap {\bf c}_3 \): all three users rendezvous and the Markov chain transitions to the final state $R_F$.
    \item \( \phi_{t+1}({\bf c}_2) \in ({\bf c}_1 \cap {\bf c}_2 \cap {\bar {\bf c}}_3) \cup({\bar {\bf c}}_1 \cap {\bf c}_2 \cap {\bf c}_3)\): user 2 meets either user 1 or user 3 and the Markov chain transitions to the mutual awareness state $R_A$.
    \item \( \phi_{t+1}({\bf c}_2) \in {\bar {\bf c}}_1 \cap {\bf c}_2 \cap {\bar {\bf c}}_3 \): no rendezvous occurs and the Markov chain remains in the same state.
\end{itemize}
\noindent
From \rlem{fairgen}, the corresponding transition probabilities are:
\begin{align}
    &\pr(X_{t+1} = R_F \mid X_t = R_{12,13}) \nonumber\\
    &= \pr(\phi_{t+1}({\bf c}_2) \in {\bf c}_1 \cap {\bf c}_2 \cap {\bf c}_3) \nonumber\\
    &= \frac{|{\bf c}_1 \cap {\bf c}_2 \cap {\bf c}_3|}{|{\bf c}_2|}
    = \frac{n_{1,2,3}}{n_2},
    \label{spread3333}
\end{align}

\begin{align}
    &\pr(X_{t+1} = R_A \mid X_t = R_{12,13})\nonumber \\
    &= \pr(\phi_{t+1}({\bf c}_2) \in ({\bf c}_1 \cap {\bf c}_2 \cap {\bar {\bf c}}_3) \cup({\bar {\bf c}}_1 \cap {\bf c}_2 \cap {\bf c}_3)) \nonumber\\
    &= \frac{|({\bf c}_1 \cap {\bf c}_2 \cap {\bar {\bf c}}_3) \cup({\bar {\bf c}}_1 \cap {\bf c}_2 \cap {\bf c}_3)|}{|{\bf c}_2|} \nonumber\\
    &= \frac{n_{1,2} - n_{1,2,3} + n_{2,3} - n_{1,2,3}}{n_2}\nonumber \\
    &= \frac{n_{1,2} + n_{2,3} - 2\cdot n_{1,2,3}}{n_2},
    \label{spread4444}
\end{align}
and
\begin{align}
    &\pr(X_{t+1} = R_{12,13} \mid X_t = R_{12,13})\nonumber \\
    &= \pr(\phi_{t+1}({\bf c}_2) \in {\bar {\bf c}}_1 \cap {\bf c}_2 \cap {\bar {\bf c}}_3) \nonumber\\
    &= \frac{|{\bar {\bf c}}_1 \cap {\bf c}_2 \cap {\bar {\bf c}}_3|}{|{\bf c}_2|} \nonumber\\
    &= \frac{n_2 - n_{1,2} - n_{2,3} + n_{1,2,3}}{n_2}.
    \label{spread5555}
\end{align}

\item Transition from the mutual awareness state

In the mutual awareness state \( R_A \), all three users have exchanged their available channel sets and thus share complete knowledge of the intersection \( {\bf c}_1 \cap {\bf c}_2 \cap {\bf c}_3 \). Each user subsequently applies the consistent channel selection function \( \phi_{t+1} \) to this common set. By \rthe{rendezvous}, all users will select the same channel at time \( t+1 \), ensuring a rendezvous. This implies that

\beq{spread6666}
\pr(X_{t+1} = R_F \mid X_t = R_A) = 1.
\eeq
\end{enumerate}

Using the relabeling of the 12 states in \rsec{spread},
the expected absorption time, denoted by \( t_i \), is then given by
\beq{spread11121}
t_i = \ex[T_i].
\eeq
From \req{spreadettr}, we have
\beq{spreadettrb}
t_i = 1 + \sum_{j =i}^{11} p_{i,j} \cdot t_j.
\eeq

Using \req{spreadettrb},
we can compute the expected absorption time from states 11 (the mutual awareness state $R_A$) back to the expected absorption time of state 1 (the initial state $R_I$). Specifically, we derive the following closed-form representation for the ETTR of the spread-out strategy in \req{spread7777} below.
\begin{itemize}
    \item Mutual awareness state (\(i = 11\)):
    \begin{align}
        t_{11} &= 1.
    \end{align}

    \item Second rendezvous states (\(i \in \{5,6,\ldots,10\}\)):
    \begin{align}
        t_i &= 1 + p_{i,i} \cdot t_i + p_{i,11} \cdot t_{11} \nonumber\\
        &= \frac{1 + p_{i,11} \cdot t_{11}}{1 - p_{i,i}}.
    \end{align}

    \item First rendezvous states (\(i \in \{2,3,4\}\)): \\
    Note that the second rendezvous states reachable from $i$ are $2i+1$ and $2i+2$.
        \begin{align}
        t_i &= 1 + p_{i,i} \cdot t_i + p_{i,2i+1} \cdot t_{2i+1} + p_{i,2i+2} \cdot t_{2i+2}\nonumber \\
        &= \frac{1}{1 - p_{i,i}} \left[ 1 + p_{i,2i+1} \cdot t_{2i+1} + p_{i,2i+2} \cdot t_{2i+2} \right].
    \end{align}

    \item Initial state (\(i = 1\)):
    \begin{align}
        t_1 &= 1 + p_{1,1} \cdot t_1 + p_{1,2} \cdot t_2 + p_{1,3} \cdot t_3 + p_{1,4} \cdot t_4 \nonumber \\
        &= \frac{1}{1 - p_{1,1}} \left[ 1 + p_{1,2} \cdot t_2 + p_{1,3} \cdot t_3 + p_{1,4} \cdot t_4 \right].
        \label{eq:spread7777}
    \end{align}
\end{itemize}

\begin{figure}[!t]
    \centering
    \includegraphics[width=2.5in]{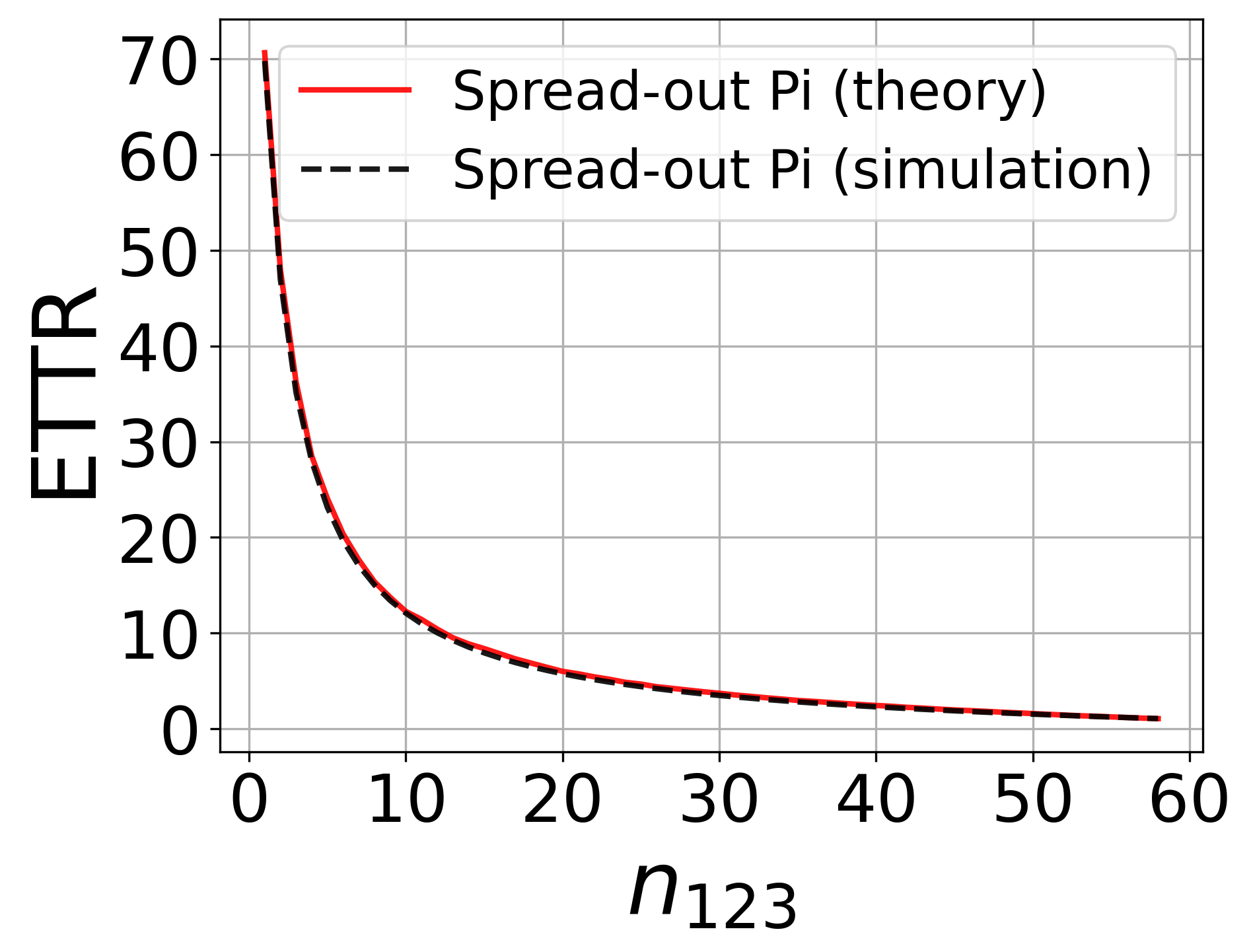}
    \caption{The ETTRs of the spread-out strategy with \( n_{123} = 1 \).}
    \label{fig:spreadout_check1}
\end{figure}

\begin{figure}[!t]
    \centering
    \includegraphics[width=2.5in]{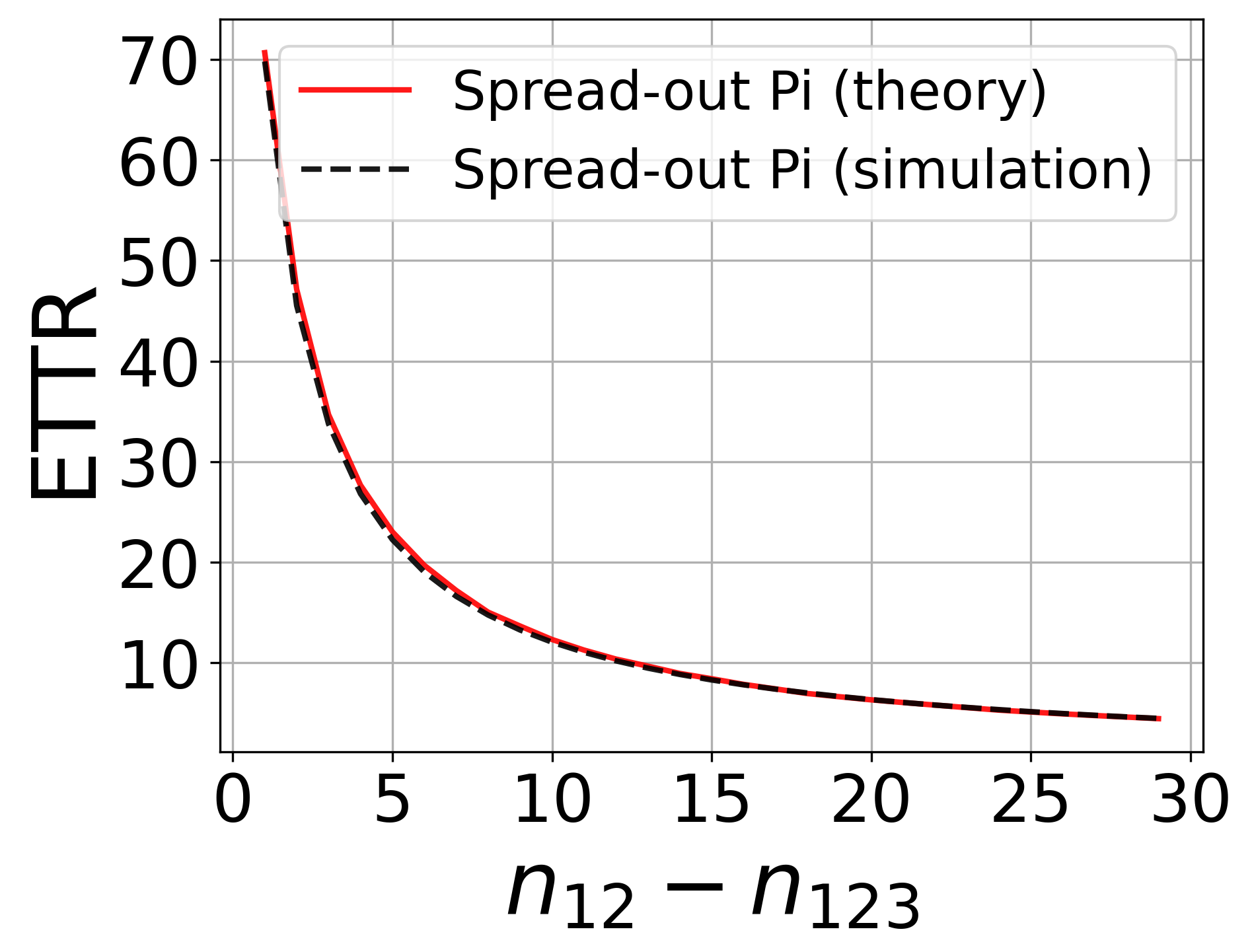}
    \caption{The ETTRs of the spread-out strategy with \( n_{\mathrm{exclusive}} = 1 \).}
    \label{fig:spreadout_check2}
\end{figure}

In \rfig{spreadout_check1} and \rfig{spreadout_check2}, we validate our Markov chain analysis with the simulation results and the theoretical results in \req{spread7777} match extremely well with the simulation results.

\vfill
\newpage

\begin{IEEEbiography}[{\includegraphics[width=1in,height=1.25in,clip,keepaspectratio]{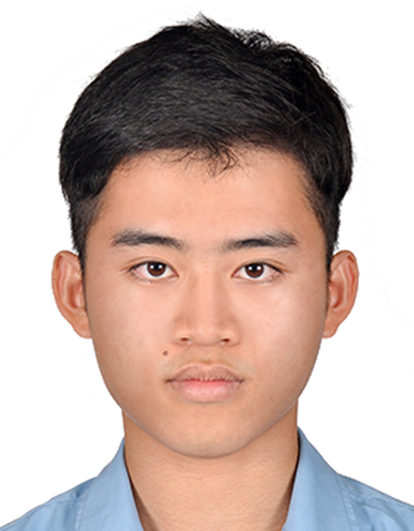}}]
{Yiwei Liu}
received a B.S. degree in Electrical Engineering in 2023 and an M.S. degree from the Institute of Communications Engineering, National Tsing Hua University, Hsinchu, Taiwan, in 2025. His research interests include the multichannel rendezvous problem and network science.
\end{IEEEbiography}

\begin{IEEEbiography}[{\includegraphics[width=1in,height=1.25in,clip,keepaspectratio]{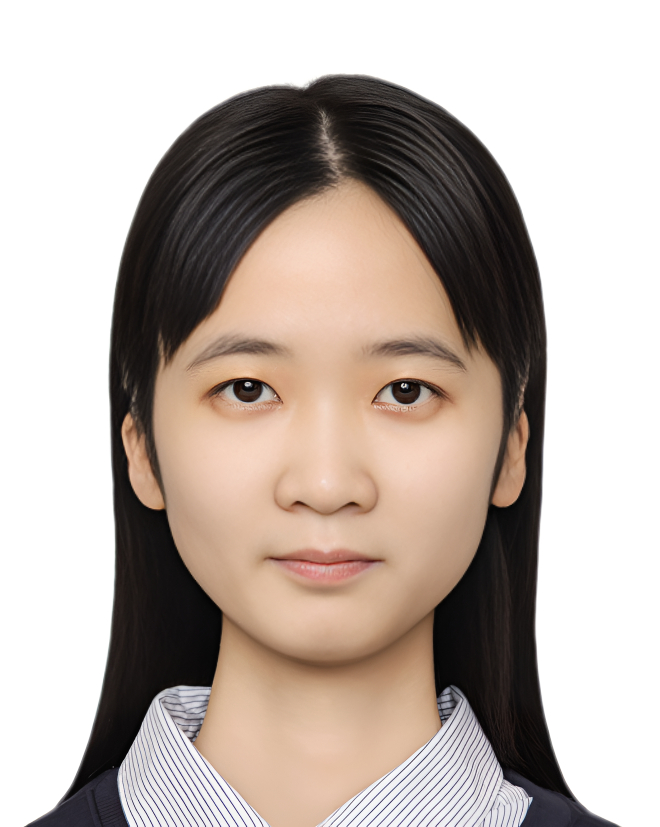}}]
{Yi-Chia Cheng}
received the B.S. degree in Communications, Navigation and Control Engineering from National Taiwan Ocean University, Keelung, Taiwan, in 2023, and the M.S. degree from the Institute of Communications Engineering, National Tsing Hua University, Hsinchu, Taiwan, in 2025. Her research interests include 5G and beyond wireless communication technologies.
\end{IEEEbiography}

\begin{IEEEbiography}[{\includegraphics[width=1in,height=1.25in,clip,keepaspectratio]{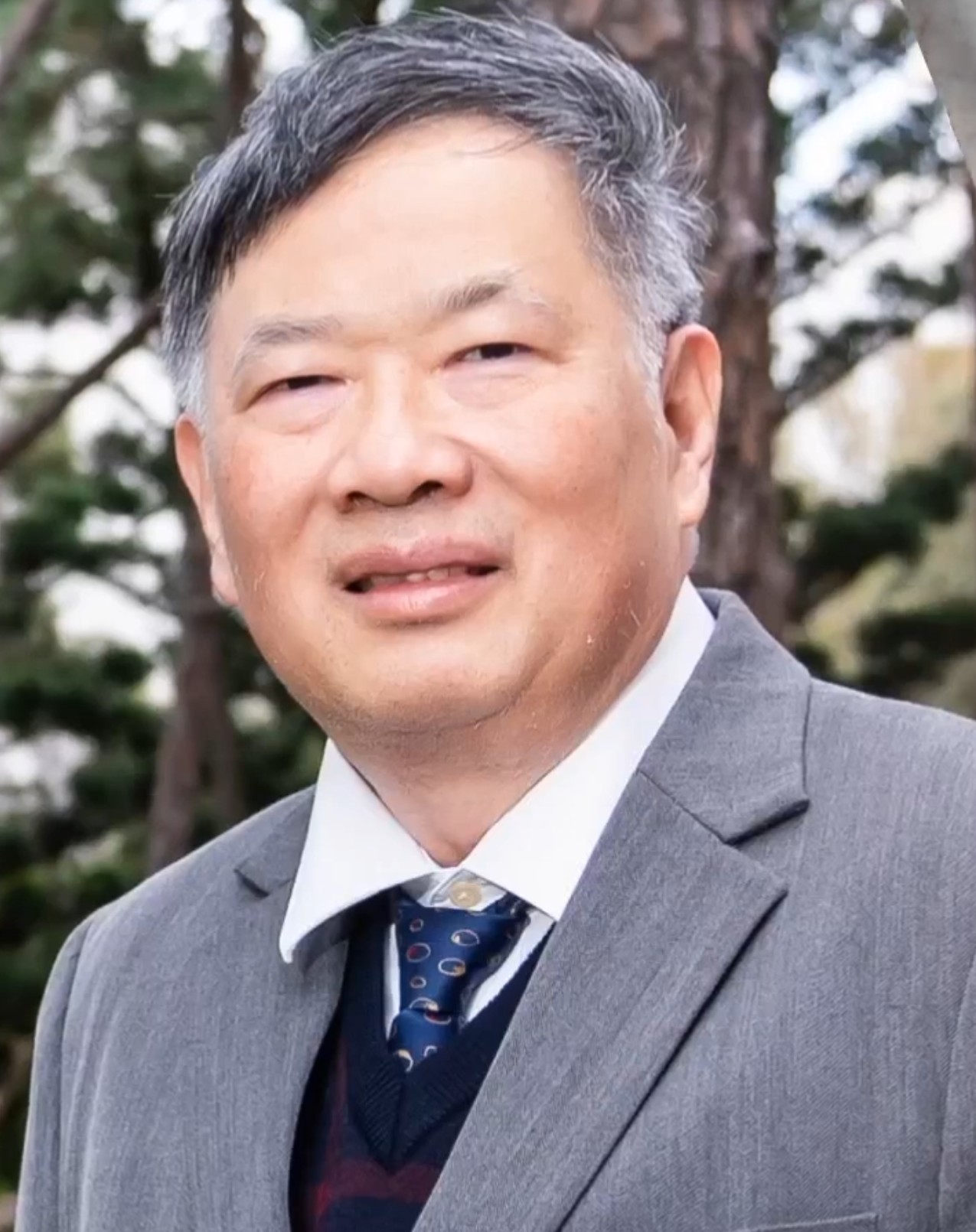}}]
	{Cheng-Shang Chang}
	(S'85-M'86-M'89-SM'93-F'04)
	received the B.S. degree from National Taiwan
	University, Taipei, Taiwan, in 1983, and the M.S.
	and Ph.D. degrees from Columbia University, New
	York, NY, USA, in 1986 and 1989, respectively, all
	in electrical engineering.
	
	From 1989 to 1993, he was employed as a
	Research Staff Member with the IBM Thomas J.
	Watson Research Center, Yorktown Heights, NY,
	USA. Since 1993, he has been with the Department
	of Electrical Engineering, National Tsing Hua
	University, Taiwan, where he is a Tsing Hua Distinguished Chair Professor. He is the author
	of the book Performance Guarantees in Communication Networks (Springer,
	2000) and the coauthor of the book Principles, Architectures and Mathematical
	Theory of High Performance Packet Switches (Ministry of Education, R.O.C.,
	2006). His current research interests are concerned with network science, big data analytics,
	mathematical modeling of the Internet, and high-speed switching.
	
	Dr. Chang served as an Editor for {\em OPERATIONS RESEARCH} from 1992 to 1999,
	an Editor for the {\em IEEE/ACM TRANSACTIONS ON NETWORKING} from 2007
	to 2009, and an Editor for the {\em IEEE TRANSACTIONS
		ON NETWORK SCIENCE AND ENGINEERING} from 2014 to 2017. He is currently serving as an Editor-at-Large for the {\em IEEE/ACM
		TRANSACTIONS ON NETWORKING}. He is a member of IFIP Working
	Group 7.3. He received an IBM Outstanding Innovation Award in 1992, an
	IBM Faculty Partnership Award in 2001, and Outstanding Research Awards
	from the National Science Council, Taiwan, in 1998, 2000, and 2002, respectively.
	He also received Outstanding Teaching Awards from both the College
	of EECS and the university itself in 2003. He was appointed as the first Y. Z.
	Hsu Scientific Chair Professor in 2002. He received the Merit NSC Research Fellow Award from the
	National Science Council, R.O.C. in 2011. He also received the Academic Award in 2011 and the National Chair Professorship in 2017 and 2023 from
	the Ministry of Education, R.O.C. He is the recipient of the 2017 IEEE INFOCOM Achievement Award.
\end{IEEEbiography}

\end{document}